\documentstyle[aps,twocolumn,eqsecnum,floats]{revtex}
\begin{document}
\draft
\title{Asymptotic space-time behavior of HTL gauge propagator}
\author{H. Arthur Weldon}
\address{Department of Physics,
West Virginia University, Morgantown, West Virginia, 26506-6315}
\date{September 20, 2000}
\maketitle
\begin{abstract}
The asymptotic behavior as $t\to\infty$ and $r\to\infty$ of the
hard-thermal-loop
propagator $^{*}\!{\cal D}^{\mu\nu}(t,\vec{r})$ is computed in the Coulomb
gauge.
The asymptotic falloff is always a power law though generally different in
the deep time-like and space-like regions.
The  contributions of  quasiparticle poles and Landau branch cuts are
computed. The most difficult calculation is the contribution of the branch
cut in the transverse propagator
$^{*}\!{\cal D}^{ij}(t,\vec{r})$. For QED this produces a leading
behavior of order
$T/r$ in both the  time-like and space-like regions. The inclusion of a
magnetic mass so as to describe QCD makes the leading behavior $1/(Tr^{3})$,
thus improving the infrared convergence.    The asymptotic space-like
behavior of all contributions (longitudinal and transverse, poles and cuts)
is  confirmed by also computing in the Euclidean formalism and analytically
continuing.
 The results are compared will those for
free gauge bosons at finite temperature.
 \end{abstract}

\section{Introduction}

The hard-thermal-loop approximation to high temperature gauge theories
developed by  Braaten and Pisarski   provides a
consistent perturbative method for calculating processes   in which one or
more of the external momenta are of order
$gT$  \cite{P1,P2,BP1,BP2,sum,JPB}. An excellent review of the entire subject
and its applications is given by Le Bellac \cite{MLB}.

For an SU(N) gauge theory  with
$N_{_f}$ flavors of massless quarks in the  fundamental representation,
in the high-temperature deconfined phase there
is an effective thermal mass for the gluon
\begin{equation}m_{g}^{2}=\Big(N+{N_{f}\over 2}\Big){g^{2}T^{2}\over
9}.\label{1a}
\end{equation}
For QED at high temperatures $\pi T\gg m_{e}$, there is an effective thermal
mass for the  photon
\begin{equation}
m_{g}=eT/3.
\end{equation}
However, these effective masses do not
screen the electric and magnetic fields produced by all time-dependent
currents. The situation is much more subtle. The phenomena of dynamical
screening applies only to currents with
frequencies below
$m_{g}$ \cite{HAW}.    When the frequency satisfies  $k_{0}<m_{g}$
then the momentum-space propagator has poles at  complex values of the
wave vector $k$. The imaginary part of $k$  produces the
screening.  In a different frequency regime, this is the reason that   AM
radio signals are exponentially attenuated in the earth's ionosphere
\cite{Mex}.
In the hard-thermal-loop approximation static magnetic fields
 with $k_{0}=0$ are not screened.
Only if
$0<k_{0}<m_{g}$ does the pole contribution to the hard-thermal-loop propagator,
$^{*}\!D^{\mu\nu\,\rm pole}(k_{0},\vec{r})$, fall  exponentially at large $r$.
However,  there are also branch cuts in the momentum-space propagator.
How the cut contribution,
$^{*}\!D^{\mu\nu\,\rm cut}(k_{0},\vec{r})$, behaves at large $r$ has not been
investigated.

Recent calculations of Boyanovsky, de Vega, et al have provided
complementary information about   the asymptotic behavior of the
hard-thermal-loop propagator at large time and fixed fixed wave-vector $k$
\cite{DB1,DB2}. Those authors were interested in non-equilibrium effects
and found
it
 convenient to use the retarded propagator,
$^{*}\!D_{R}^{\mu\nu}(t,\vec{k})$. Being the Fourier transform of the fixed
 frequency propagator, it necessarily
contains the effects of all frequencies.
They discovered that the  Landau branch cuts are extremely important  for the
asymptotic behavior of
$^{*}\!D_{R}^{\mu\nu}(t,\vec{k})$ and found power-law falloff rather
than exponential.    Similar results for
current-current correlators were obtained by by Arnold and Yaffe \cite{Y1}.

This paper will examine the  asymptotic behavior in both time  and space
 of the  hard-thermal-loop propagator
$^{*}D^{\mu\nu}(t,\vec{r})$.  The propagator considered will be in the Coulomb
gauge and it will be the time-ordered propagator, not the retarded.
  The asymptotic behavior in the limit $t\to\infty$, $r\to\infty$ with $r/t$
fixed is controlled
by the region in which the energy $k_{0}$ and momentum $k$ are both of
order $m_{g}$,
which is precisely the region in which the momentum-space propagator
 differs significantly  from the free propagator.
Reaching the asymptotic regime will generally require
$t\gg 1/m_{g}$ and $r\gg 1/m_{g}$.  The region far from the space-time origin
with $r/t<1$ will be called the deep time-like region and that with
$r/t>1$ will be called the deep space-like region.

It is  difficult to anticipate the asymptotic behavior.
The case of a free, massive vector boson is an interesting comparison.
In the deep time-like region that propagator falls like a power, $1/t^{3/2}$;
in the deep space-like region it falls exponentially, $(T/r)\exp(-Mr)$.

For the hard-thermal-loop propagator the one might expect the quasiparticle
poles to produce an asymptotic dependence like that of a massive
boson. In the the deep time-like region this is correct, i.e. $1/t^{3/2}$.
However,
in  the deep space-like region  the quasiparticle pole contributions to
both the
longitudinal and transverse propagators produces power-law  fall-off.   The
failure of the quasiparticle poles to produce an exponential fall-off confirms
that the electric and magnetic fields produced by localized, time-dependent
currents are not generally screened since those sources can contain
frequencies higher than $m_{g}$.

It is even more difficult to anticipate the
effect of the branch cuts in the hard thermal loop propagator since free
propagators can offer no guide.
The most surprising result is the effect produced by the branch cut in the
transverse propagator.  The asymptotic behavior in either the deep space-like
or time-like regions is the same:
\begin{equation}
^{*}{\cal D}^{ij\,\rm cut}(t,\vec{r})\to
{-iT\over 8\pi r}\big(\delta^{ij}+\hat{r}^{i}\hat{r}^{j}\big)
+\dots.\label{13}
\end{equation}
In the deep space-like region the $T/r$ behavior is not surprising
since that is the leading behavior of  free thermal propagators.
However  in the deep time-like region  the $T/r$ behavior is very non-trivial.
In contrast, free propagators   in the
deep time-like region fall exponentially \cite{Y1,AW}.

In QCD the inclusion of a magnetic screening mass will change the
 above effect of the  branch cut in the  transverse
propagator. This change will be calculated. In QED there is no magnetic mass
and Eq. (\ref{13}) applies.

The paper develops systematically  the mathematical machinery necessary for
the analysis. There are, of necessity, many contributions to evaluate. There
are two propagators, longitudinal and transverse. Each propagator has a pole
and a cut to consider. Each contribution must be evaluated in the time-like
and in the space-like limits.  The emphasis throughout is on extracting the
asymptotic power-law behavior of the propagators. Exponentially small
corrections are always omitted.
 Section II examines the simple example of the free photon propagator  in
Feynman gauge.  Section III summarizes how gauge boson propagators, free or
interacting, are expressed in terms of spectral functions. The real subject of
the paper, the asymptotic behavior of the hard-thermal-loop propagator, is
confronted in Secs. IV and V  which treat the longitudinal  and the
transverse propagators, respectively.   Section VI provides a  detailed
summary of the results. For
many readers it may be best to read  the results in Sec. VI before reading
 how they were calculated.

There are four appendices.  Appendix A displays the hard-thermal-loop spectral
functions in other gauges.  Appendix B proves the lemma that is used in the
text
to determine the asymptotic behavior in the deep space-like region.
Appendix C computes the
asymptotic behavior of $^{*}{\cal D}^{\mu\nu}(t,\vec{r})$ in the deep
space-like region by using the Euclidean time formalism.  The sum of  the
static
and non-static contributions agree exactly with the sum of the pole and cut
contributions that are  computed directly in Minkowski space in Secs. IV  and
V. Appendix D contains details of the most difficult calculation, the
asymptotic
behavior of $^{*}{\cal D}^{ij\,\rm cut}(t,\vec{r})$ in the deep time-like
region.

\section{Simple example}

Before confronting the problem of hard-thermal-loop propagators it is
useful to examine the much simpler problem of thermal propagators for free
gauge
bosons, i.e. those without hard-thermal-loop resummation. The reason for
this example is to illustrate two points that will be
important later. However, the free thermal propagators are important in their
own right in two different contexts: (i)  In the low temperature regime of QED,
$\pi T\ll m_{e}$, no resummation is necessary for the photon
propagator. (ii) In the high temperature regime of
QED or QCD the ghost propagators do not require resummation and the free
thermal form applies \cite{BP1}.

The free propagator is simple to compute in the Feynman
gauge and provides two useful comparisons with the hard-thermal-loop results
that will come later.
In the Euclidean time formalism the Feynman gauge propagator is
\begin{equation}
{\cal D}_{E}^{\mu\nu}(\tau,\vec{r})
=g^{\mu\nu}iT\!\!\sum_{n=-\infty}^{\infty}\!
\int\!\!{d^{3}k\over
(2\pi)^{3}}\,{e^{i\vec{k}\cdot\vec{r}-i\omega_{n}\tau}
\over \omega_{n}^{2}+k^{2}},\label{21}
\end{equation}
where $\omega_{n}=2\pi nT$ and $-\beta\le\tau\le\beta$. The poles at $k=\pm
i\omega_{n}$ allow the  momentum integration to be done by Cauchy's theorem:
\begin{displaymath}
{\cal D}^{\mu\nu}_{E}(\tau,\vec{r})=g^{\mu\nu}{iT\over 4\pi r}\bigg[
1+2\sum_{n=1}^{\infty}\cos(\omega_{n}\tau)e^{-\omega_{n}r}
\bigg].\end{displaymath}
The time-ordered propagator in real time  is obtained by the replacement
$\tau\to it$:
\begin{displaymath}
{\cal D}^{\mu\nu}(t,\vec{r})=g^{\mu\nu}{iT\over 4\pi r}\bigg[
1+2\sum_{n=1}^{\infty}\cosh(\omega_{n} t)e^{-\omega_{n}r}
\bigg]\end{displaymath}
 In the space-like region
$r\!>\!t$ each term in the sum is exponentially small. Therefore the
asymptotic behavior in the deep space-like region is
$iT/4\pi r$.  However, in the time-like region $t\!>\!r$ the terms in the
series
grow exponentially with $n$. To evaluate the behavior in the time-like
region it is necessary to first sum the series in the space-like region (the
region of convergence)  and then continue that finite  result to the time-like
domain. Since the series is geometrical, the summation is elementary and gives
\begin{displaymath}
{\cal D}^{\mu\nu}_{>}(t,\vec{r})=g^{\mu\nu}{iT\over 4\pi r}\bigg[
1+{1\over e^{2\pi T(r-t)}-1}+{1\over e^{2\pi T(r+t)}-1}\bigg].
\end{displaymath}
When $r\to\infty$ this still behaves as $1/r$ plus exponentially small
corrections. The surprise is that when $t\to\infty$ this is exponentially
small. More generally, for $r\to\infty$ and $t\to\infty$
with a fixed ratio $r/t<1$, the propagator is exponentially small.

 There are two lessons to be drawn. (i) The behavior in
the deep space-like region can be  easily calculated using the
Euclidean formalism.
(ii) The behavior in the deep time-like region can be quite  different
from the deep space-like.
Unless one is able to perform the sum over frequencies, the Euclidean
propagator cannot provide information about the asymptotic behavior of the
Minkowski propagator in the deep time-like region.

This example is misleading in one respect.
For the free  propagator the leading space-like behavior, $T/r$,  comes
entirely from the static $n=0$ mode.  For the
hard-thermal-loop propagators the leading behavior of
$^{*}{\cal D}^{00}(t,r)$ will  come from the non-static modes.

\section{General structure}

Before getting to the HTL propagators,
it is useful to display the underlying structure of the gauge boson
propagator that holds whatever the approximation.
The basic two-point functions or thermal Wightman functions are
\begin{eqnarray}
{\cal D}^{\mu\nu}_{>}(x)=&&-i{\rm Tr}\big[\varrho\, A^{\mu}(x)
A^{\nu}(0)\big]\nonumber\\
{\cal D}^{\mu\nu}_{<}(x)=&&-i{\rm Tr}\big[\varrho\, A^{\nu}(0) A^{\mu}(x)\big],
\nonumber\end{eqnarray}
where $\varrho=\exp(-\beta H)/{\rm Tr}[\exp(-\beta H)]$ is the
 density operator at temperature $T=1/\beta$.

\paragraph*{Time-ordered propagator.} The propagator that will  be
calculated later is the  time-ordered propagator:
\begin{equation}
{\cal D}^{\mu\nu}(t,\vec{r})=\theta(t)\,{\cal D}^{\mu\nu}_{>}(t,\vec{r})
+\theta(-t)\,{\cal D}^{\mu\nu}_{<}(t,\vec{r}).\label{3a}
\end{equation}
What is actually known from   hard-thermal-loop calculations is
 the  propagator in momentum space or equivalently the spectral function
since the time-ordered propagator  can be expressed as
\begin{equation}
D^{\mu\nu}(k_{0},\vec{k})=\int_{-\infty}^{\infty}\!{d\sigma \over 2\pi}
\,{\rho^{\mu\nu}(\sigma,\vec{k})\over
k_{0}\!-\!\sigma+i\epsilon}-i\,{\rho^{\mu\nu}(k_{0},\vec{k})\over e^{\beta
k_{0}}-1}.\label{2prop}
\end{equation}
Knowledge of  the momentum-space propagator immediately yields the
 spectral function
since
\begin{displaymath}
{\rm Im}\, D^{\mu\nu}(k_{0},\vec{k})=-\rho^{\mu\nu}(k_{0},\vec{k})\bigg[
{1\over 2}+{1\over e^{\beta k_{0}}-1}\bigg].
\end{displaymath}
 The Fourier transform of Eq. (\ref{3a}) implies that the two-point functions
in momentum space are
\begin{eqnarray}
D_{>}^{\mu\nu}(k_{0},\vec{k})=&&-i\rho^{\mu\nu}(k_{0},\vec{k})
\bigg[1+{1\over e^{\beta k_{0}}-1}\bigg]\nonumber\\
D_{<}^{\mu\nu}(k_{0},\vec{k})=&&-i\rho^{\mu\nu}(k_{0},\vec{k})
\;{1\over e^{\beta k_{0}}-1}.\nonumber
\end{eqnarray}
Knowing the spectral function  allows one to calculate the two-point
functions in space-time by Fourier transforming:
\begin{eqnarray}
{\cal D}_{>}^{\mu\nu}(t,\vec{r})=&&-i\int\!{d^{4}k\over
(2\pi)^{4}}\,e^{i\vec{k}\cdot\vec{r}-ik_{0}t}\;{\rho^{\mu\nu}(k_{0},\vec{k})
\over 1-e^{-\beta k_{0}}}\label{3c}\\
{\cal D}_{<}^{\mu\nu}(t,\vec{r})=&&-i\int\!{d^{4}k\over
(2\pi)^{4}}\,e^{i\vec{k}\cdot\vec{r}-ik_{0}t}\;{\rho^{\mu\nu}(k_{0},\vec{k})
\over e^{\beta k_{0}}-1}.\nonumber
 \end{eqnarray}
The asymptotic behavior of this Fourier transform is the subject of this
paper. For the asymptotic behavior the time will always be positive (and
large) so that the time-ordered propagator is the same as the thermal
Wightman function.

\paragraph*{Retarded propagator.} For certain applications the retarded
propagator is of interest \cite{DB1,DB2}. In space-time the retarded
propagator is
\begin{equation}
{\cal D}^{\mu\nu}_{R}(t,\vec{r})=\theta(t)\,\big[{\cal
D}^{\mu\nu}_{>}(t,\vec{r})-{\cal
D}^{\mu\nu}_{<}(t,\vec{r})\big].\label{3d}
\end{equation}
In momentum space this becomes
\begin{displaymath}
D_{R}^{\mu\nu}(k_{0},\vec{k})=\int_{-\infty}^{\infty}\!{d\sigma
\over 2\pi}\,{\rho^{\mu\nu}(\sigma,\vec{k})\over k_{0}-\sigma+i\epsilon}.
\end{displaymath}
Knowing the spectral function  allows computation of the
space-time dependence by
\begin{equation}
{\cal D}_{R}^{\mu\nu}(t,\vec{r})=-i\int\!{d^{4}k\over
(2\pi)^{4}}\,e^{i\vec{k}\cdot\vec{r}-ik_{0}t}\;\rho^{\mu\nu}(k_{0},\vec{k}).
\label{3e}\end{equation}
Comparison of Eq. (\ref{3c}) with Eq. (\ref{3e}) shows that the
time-ordered propagator contains the
Bose-Einstein function but the retarded propagator does not.
As $t\to\infty$ the time-ordered propagator is more sensitive to the low
frequency part of the spectral function.

\paragraph*{Imaginary time propagator.} The imaginary-time, or Euclidean,
propagator already used in Sec. II is ordered in the variable $\tau$:
\begin{displaymath}
{\cal D}_{E}^{\mu\nu}(\tau,\vec{r})=\theta(\tau)
{\cal D}_{>}^{\mu\nu}(-i\tau,\vec{r})
+\theta(-\tau)
{\cal D}_{<}^{\mu\nu}(-i\tau,\vec{r}),
\end{displaymath}
 where $-\beta\le\tau\le\beta$. If one computes the Euclidean propagator
for $0\le \tau\le\beta$ and then analytically continues  $\tau$ to
the imaginary axis ($\tau\to it$), the result will be ${\cal
D}_{>}^{\mu\nu}(t,\vec{r})$. Thus analytic continuations in time relates the
Euclidean propagator to the time-ordered propagator as used in
sample calculation of Sec. II.
By contrast, if  one knows the Euclidean propagator as a function of
frequency,
$D^{\mu\nu}_{E}(\omega_{n},\vec{r})$, then the continuation  $\omega_{n}\to
ik_{0}-\epsilon$ will give the retarded propagator
$D^{\mu}_{R}(k_{0},\vec{r})$.

\paragraph*{HTL spectral functions in Coulomb gauge.} Calculations will be
done in the Coulomb gauge, which is used extensively in applications of
hard-thermal-loops
\cite{MLB}.  In this gauge the spectral
function has the tensor structure
\begin{eqnarray}
\rho^{00}(k_{0},\vec{k})=&&\rho_{\ell}(k_{0},k)\nonumber\\
\rho^{ij}(k_{0},k)=&&(\delta^{ij}-\hat{k}^{i}\hat{k}^{j})\,\rho_{t}(k_{0},k),
\label{spectral}\end{eqnarray}
with six  components vanishing: $\rho^{0j}=\rho^{j0}=0$.
The longitudinal and transverse spectral functions, introduced by Pisarski
\cite{P1}, both have the form
\begin{eqnarray}
{1\over 2\pi}\,\rho_{s}(k_{0},k)=&&
Z_{s}\big[\delta(k_{0}-\omega_{s})-\delta(k_{0}+\omega_{s})\big]\nonumber\\
+&& \theta(k^{2}-k_{0}^{2})\,\beta_{s}(k_{0},k),
\end{eqnarray}
where $s$ denotes either $\ell$ or $t$. Both $\rho_{\ell}$ and $\rho_{t}$
are odd functions of $k_{0}$. The Dirac delta functions come from the pole
in the
momentum-space propagator at the quasiparticle energies $\pm\omega_{s}(k)$. The
residues of those poles are the real functions $Z_{s}(k)$. The continuum
contribution
$\beta_{s}$ come from the discontinuity across the Landau branch cut
in the momentum space propagators.

\section{Longitudinal propagator: $^{*}{\cal D}^{00}(\lowercase{x})$}

The main subject is the computation of the asymptotic behavior of the
hard-thermal-loop, time-ordered propagator $^{*}{\cal
D}^{\mu\nu}(t,\vec{r})$ which  for  positive $t$ is the same as the
Wightman function
$^{*}{\cal D}_{>}^{\mu\nu}(t,\vec{r})$.
This section treats
$^{*}{\cal D}_{>}^{00}(t,\vec{r})$, the Wightman function for gauge bosons
with time-like polarizations in the Coulomb gauge.
In momentum space it is determined by the longitudinal spectral function:
\begin{equation}
^{*}D_{>}^{00}(k_{0},k)=-i\,{\rho_{\ell}(k_{0},k)\over 1-e^{-\beta k_{0}}}.
\label{31}\end{equation}
The spectral function is
\begin{eqnarray}
{1\over 2\pi}\,\rho_{\ell}(k_{0},k)=&&
Z_{\ell}\big[\delta(k_{0}-\omega_{\ell})
-\delta(k_{0}+\omega_{\ell})\big]\nonumber\\
+&&\theta(k^{2}-k_{0}^{2})\,\beta_{\ell}(k_{0},k),
\nonumber\end{eqnarray}
where the residue function is \cite{P1}
\begin{displaymath}
Z_{\ell}={\omega_{\ell}(\omega_{\ell}^{2}-k^{2})\over
k^{2}(3m_{g}^{2}-\omega_{\ell}^{2}+k^{2})},
\end{displaymath}
and $m_{g}$ is the thermal gluon mass.
The longitudinal  spectral function for the cut is \cite{P1}
\begin{equation}
\beta_{\ell}(k_{0},k)={2\over 3m_{g}^{2}}{x\over N_{\ell}(x,k)},
\label{32}\end{equation}
where $x=k_{0}/k$ and the denominator is
\begin{equation}
N_{\ell}(x,k)=\Big[{2k^{2}\over 3m_{g}^{2}}\!+\!2\!-\!
x\,\ln\Big({1+x\over 1-x}\Big)\Big]^{2}+\big[\pi x\big]^{2}
.\label{33}\end{equation}
The longitudinal spectral function satisfies various sum rules
\cite{sum,MLB}. The one that will be useful subsequently is
\begin{equation}
\int_{-k}^{k}\!dk_{0}\;{\beta_{\ell}
(k_{0},k)\over k_{0}}={3m_{g}^{2}\over
k^{2}(3m_{g}^{2}+k^{2})}-{2Z_{\ell}\over\omega_{\ell}}.
\label{34}\end{equation}
To compute the  Fourier transform of Eq. (\ref{31}) it is efficient to
organize the calculation into two steps:
\begin{eqnarray}
^{*}{\cal D}_{>}^{00}(t,r)=&&-i\int\!{d^{3}k\over
(2\pi)^{3}}\;e^{i\vec{k}\cdot\vec{r}} F_{\ell}(t,k),\nonumber\\
F_{\ell}(t,k)=&&\int_{-\infty}^{\infty}\!{dk_{0}\over 2\pi}{e^{-ik_{0}t}\over
1-e^{-\beta k_{0}}}\,\rho_{\ell}(k_{0},k).\nonumber
\end{eqnarray}
The integral over the solid angles of $\hat{k}$  can be done
directly:
\begin{equation}
^{*}{\cal D}_{>}^{00}(t,r)={-i\over
2\pi^{2}r}\int_{0}^{\infty}\!dk\,k\sin(kr) F_{\ell}(t,k).\label{35}
\end{equation}
The function $F_{\ell}$ is a sum of the pole and cut contributions
$F_{\ell}=F_{\ell}^{\rm pole}+F_{\ell}^{\rm cut}$, where
\begin{eqnarray}
F_{\ell}^{\rm pole}(t,k)=&&Z_{\ell}\bigg[{e^{-i\omega_{\ell}t}
\over 1-e^{-\beta\omega_{\ell}}}+{e^{i\omega_{\ell}t}\over
e^{\beta\omega_{\ell}}-1 }\bigg],\label{36}\\
F_{\ell}^{\rm cut}(t,k)=&&\int_{-k}^{k}\!dk_{0}
\,{e^{-ik_{0}t}\over 1-e^{-\beta k_{0}}}\,\beta_{\ell}(k_{0},k).
\label{37}
\end{eqnarray}
The Wightman function at fixed three-momentum is $^{*}\!{\cal
D}^{00}_{>}(t,k) =-iF_{\ell}(t,k)$.

\subsection{Asymptotic time-like behavior of $\;^{*}{\cal D}^{00}_{>}(x)$}

The deep time-like region requires   $t\to \infty$ and
$r\to\infty$ with a fixed ratio $t/r>1$.

\subsubsection{ Pole contribution.}
The contribution of the
quasiparticle pole is
 \begin{eqnarray}
^{*}{\cal D}_{>}^{00\,\rm pole}(t,r)=&&{-i\over 2\pi^{2}r}
\int_{0}^{\infty}\!dk\;
k\,\sin kr\nonumber\\
&&\times Z_{\ell}\bigg[{e^{-i\omega_{\ell}t}\over
1-e^{-\beta \omega_{\ell}}}-{e^{i\omega_{\ell}t}\over
1-e^{\beta \omega_{\ell}}}\bigg].\nonumber
\end{eqnarray}
 Extending the integration to negative $k$ gives
 \begin{eqnarray}
^{*}{\cal D}_{>}^{00\,\rm pole}(t,r)={-1\over
4\pi^{2}r}&&\int_{-\infty}^{\infty}\!dk\nonumber\\
&&\times\, k
Z_{t}\bigg[{e^{ikr-i\omega_{t}t}\over 1-e^{-\beta
\omega_{t}}}+{e^{-ikr+i\omega_{t}t}\over  1-e^{\beta
\omega_{t}}}\bigg].\nonumber
\end{eqnarray}
The asymptotic behavior  of
this will be computed by the method of stationary phase
\cite{cop,din}. Define
\begin{equation}
\phi(k)=kr-\omega_{\ell}(k) t.\label{lphase}\end{equation}
At large values of $r$ and $t$ the functions $e^{\pm i\phi}$ oscillate
rapidly except at special values of $k$.  Let $\overline{k}$ be the value of
the wave vector at which the phase is stationary:
\begin{displaymath}
0=\bigg[{d\phi\over dk}\bigg]_{\overline{k}}=r-\bigg[{d\omega_{\ell}\over dk}
\bigg]_{\overline{k}}\;t .\end{displaymath}
This requires that the group velocity of the quasiparticle satisfy
\begin{equation}
\bigg[{d\omega_{\ell}\over dk}
\bigg]_{\overline{k}}={r\over t}< 1.
\end{equation}
The group velocity of the longitudinal dispersion relation is zero at $k=0$ and
grows monotonically with $k$, approaching 1 as $k\to\infty$ \cite{MLB}.
Therefore there is  always a real
$\overline{k}$  which satisfies this.  Near the stationary point the phase is
\begin{displaymath}
\phi(k)=\phi(\overline{k})-{1\over
2}(k-\overline{k})^{2}\,t\,\bigg[{d^{2}\omega_{\ell}\over
dk^{2}}\bigg]_{\overline{k}}+\dots.
\end{displaymath}
The acceleration $d^{2}\omega_{\ell}/dk^{2}$ is always positive.
To evaluate the contribution of the stationary phase point, the $k$ contour
must be
deformed into the complex plane so that it passes through the real value of
$\overline{k}$ on a path of steepest descent. For the first integrand,
$\exp(+i\phi)$, this requires $k=\overline{k}+e^{-i\pi/4}s$
where $s$ is real and gives a simple Gaussian integral:
\begin{displaymath}
\int dk\,e^{i\phi}= e^{i\overline{\phi}-i\pi/4}\int ds\,e^{-s^{2}t
\,\overline{\omega}''/2}.
\end{displaymath}
For the second integrand, $\exp(-i\phi)$, the correct path is
$k=\overline{k}+e^{i\pi/4}s$. Performing the two Gaussian integrations gives
\begin{eqnarray}
^{*}{\cal D}^{00\,\rm pole}(t,r)\to {-\overline{k}\;\overline{Z}_{\ell}
\over r(2\pi)^{3/2}}&&{1\over\sqrt{t\,\overline{\omega}_{\ell}''}}\label{4c}\\
\times\bigg[&&{e^{i\overline{\phi}-i\pi/4}\over
1-e^{-\beta\overline{\omega}_{\ell}}} -{e^{-i\overline{\phi}+i\pi/4}\over
e^{\beta\overline{\omega}_{\ell}}-1}
\bigg].\nonumber
\end{eqnarray}
Since $r/t$ is fixed, the value of $\overline{k}$ does not change
while
$t\to\infty$. The dependence on the size of $t$
 arises only from the power behavior  $1/(rt^{1/2})$ and from
the phase factor $\overline{\phi}=\overline{k}r-\omega_{\ell}(\overline{k})t$.

\subsubsection{Cut contribution} To evaluate the cut contribution in the
deep time-like region it
is convenient to replace the integration over $k_{0}$ by an integration
over the
dimensionless ratio $x=k_{0}/k$ so that
\begin{eqnarray}
^{*}{\cal D}_{>}^{00\,\rm
cut}(t,r)\!=&&\!{-1\over (2\pi)^{2}r}\!\int_{-\infty}^{\infty}\! dk
\!\int_{-1}^{1}\!dx\,e^{ik(r-xt)}I_{\ell}(x,k)\nonumber\\
I_{\ell}(x,k)=&&{k^{2}\beta_{\ell}(kx,k)\over 1-e^{-\beta xk}}.
\label{3A4}\end{eqnarray}
The phase  $\phi(x,k)=k(r-xt)$ has
vanishing first derivatives with respect to $x$ and $k$ at the
 stationary point
$\overline{x}=r/t$,
$\overline{k}=0$. In order to perform a Gaussian integration about
 this stationary
point it is convenient to change to new  variables $u$ and $\theta$
 defined by
\begin{mathletters}\begin{eqnarray}
x={r\over t}+&&{1\over(m_{g}t)^{1/2}}\;u e^{i\theta-i\pi/4}\\
 k=&&\Big({m_{g}\over
t}\Big)^{1/2}\;u\, e^{-i\theta-i\pi/4},
\end{eqnarray}\label{uv}\end{mathletters}
where $u$ is positive and $\theta$ is between $0$ and $2\pi$.
 We are concerned with
the limit $m_{g}t\to\infty$. As long as
$u\ll(m_{g}t)^{1/2}$ then the variables $x$ and $k$ are only  slightly
extended into the complex plane.
With this substitution the complex phase factor becomes   a
 Gaussian:
\begin{displaymath}
e^{ik(r-xt)}=e^{-u^{2}}.
\end{displaymath}
This is exact.
 The Jacobian of the  transformation gives
$dk dx=u du\, d\theta\, (2/t)$.
The cut contribution  becomes
\begin{displaymath}
^{*}{\cal D}_{>}^{00\,\rm cut}(t,r)={-1\over 2\pi^{2}rt}\int_{0}^{u_{\rm
max}}\!
udu
\int_{0}^{2\pi}\!d\theta\;
e^{-u^{2}}I_{\ell}(x,k).
\end{displaymath}
Because of the Gaussian,  values of $u$ that are much larger than 1 make
negligible contribution so that the upper limit $u_{\rm max}$  can
effectively be taken as
$\infty$. The variable change in Eq. (\ref{uv}) guarantees
that $x\to \overline{x}=r/t$ and $k\to 0$ in the limit $m_{g}t\gg 1$.
(Note that  the function $N_{\ell}(x,k)$ contains momentum via the ratio
$k/m_{g}\sim(m_{g}t)^{1/2}
\to 0$.)

The
leading term in the integration would normally come from the value of
$I_{\ell}(x,k)$ at the stationary point. However  $I_{\ell}(x,k)$ vanishes at
$k=0$. To find the first non-vanishing contribution it is necessary to expand
$I_{\ell}(x,k)$ in a double Taylor series:
\begin{displaymath}
I_{\ell}(x,k)=\sum_{m=0}^{\infty}\sum_{n=0}^{\infty}{(x-\overline{x})^{m}
(k)^{n}\over
m!n!}\bigg[{\partial^{m+n}I_{\ell}\over
\partial x^{m}\partial k^{n}}\bigg]_{\overline{x},\overline{k}}.
\end{displaymath}
The same type of series expansion will be used in Sec. V A for the transverse
propagator. It is therefore worthwhile to analyze the
implications  of this
expansion without   using the fact that for the longitudinal case
$I_{\ell}(x,k)$
vanishes at $k=0$.

The first general observation  is that because of the variable change
in Eq. (\ref{uv}),
\begin{displaymath}
(x-\overline{x})^{m}(k)^{n}=\Big({ue^{-i\pi/4}\over\sqrt{t}}\Big)^{m+n}
\Big({e^{i\theta}\over\sqrt{m_{g}}}\Big)^{m-n}.
\end{displaymath}
The integral over $\theta$ will vanish unless $m=n$. Therefore the only
parts of the
Taylor series expansion that will survive the $\theta$ integration are
\begin{displaymath}
I_{\ell}(x,k)\to\sum_{n=0}^{\infty}{1\over
n!n!}\Big({u^{2}e^{-i\pi/2}\over t}\Big)^{n}
\bigg[{\partial^{2n}I_{\ell}\over
\partial x^{n}\partial
k^{n}}\bigg]_{\overline{x},\overline{k}}.
\end{displaymath}
 The first few
terms in the asymptotic series are
\begin{eqnarray}
\int_{0}^{\infty}\!\!udu &&\int_{0}^{2\pi}\!\!d\theta\,
e^{-u^{2}}I_{\ell}(x,k) \label{3A6}\\
&&\to \pi I_{\ell}(\overline{x},0)-{i\pi\over t}
\Big[{\partial^{2} I_{\ell}\over\partial x\partial k}\Big]_{
\overline{x},0}
-{\pi\over 2t^{2}}\Big[{\partial^{4} I_{\ell}\over\partial^{2} x\partial^{2}
k}\Big]_{
\overline{x},0}.\nonumber
\end{eqnarray}
with the next correction of order $1/t^{3}$.

For the longitudinal propagator the first term vanishes:
$I_{\ell}(\overline{x},0)=0$
as mentioned.  The necessary second derivative is
\begin{displaymath}
\Big[{\partial^{2} I_{\ell}\over\partial x\partial k}\Big]_{
\overline{x},0}={2T\over 3m_{g}^{2}}\bigg[{\partial \over\partial
\overline{x}}{1\over N_{\ell}(\overline{x},0)}\bigg].
\end{displaymath}
  Therefore the asymptotic behavior of the cut
contribution in the deep time-like region for $m_{g}t\gg 1$  is
\begin{equation}
^{*}{\cal D}^{00\,\rm cut}(t,r)\to {iT\over 3\pi m_{g}^{2}\,rt^{2}}
\bigg[{\partial \over\partial \overline{x}}{1\over
N_{\ell}(\overline{x},0)}\bigg].\label{3Alast}
\end{equation}
The quantity in square brackets is  rather complicated, but it depends only
on the
ratio
$\overline{x}=r/t$. Thus the asymptotic behavior is entirely given by the
explicit
powers, $1/(rt^{2})$. This is a more rapid fall-off than $1/(rt^{1/2})$,
the pole
contribution given in Eq. (\ref{4c}).

\subsection{Asymptotic space-like behavior of $\;^{*}{\cal D}^{00}(x)$}

The limit $t\to\infty$ and
$r\to\infty$ with  fixed ratio $t/r< 1$ is the deep space-like region.
As is obvious from the analysis of the pole and cut contributions
in Sec.  A,  when $r/t> 1$ there will be no points of stationary phase
for real values of $k$ and $k_{0}$. Stationary phase points at complex values
of $k$ or $k_{0}$ will produce terms that fall exponentially.
The largest effects will instead be endpoint contributions that come from the
region of small $k$ in the integral
\begin{equation}
^{*}{\cal D}_{>}^{00}(t,r)={-i\over
2\pi^{2}r}\int_{0}^{\infty}\!dk\,k\sin(kr) F_{\ell}(t,k).\label{3B1}
\end{equation}
The endpoint contributions are evaluated using the lemma proven in Appendix
\ref{sec:lemma},
which depends upon $F_{\ell}(t,k)$ being an even function of $k$ that is
infinitely differentiable on the the real $k$ axis. Using repeated integration
by parts it is possible to show that if
$F_{\ell}(t,k)$ is finite at $k=0$ then the integral falls exponentially as
$r\to\infty$. If however $F_{\ell}(t,k)\to c_{0}/k^{2}$ with $c_{0}$ a constant
as
$k\to 0$ then the integral is of order $1/r$ with exponentially small
corrections. Specifically
\begin{equation}
\lim_{r\to\infty}\int_{0}^{\infty}\!\!dk \,k\sin(kr)\,F_{\ell}(t,k)
={c_{0}\pi\over 2r}+{\rm exp\; small}.\label{3B2}\end{equation}
This will be used  here and again in Sec. V B.

\subsubsection{Pole contribution}

 The behavior of $F_{\ell}^{\rm pole}(t,r)$
shown in Eq. (\ref{36}) at small $k$ depends upon the behavior of
the quasiparticle residue $Z_{\ell}$. In the low momentum limit
$\omega_{\ell}\to m_{g}$
and  $Z_{\ell}\to m_{g}/2k^{2}$ so that
\begin{displaymath}
k\to 0:\hskip0.5cm F_{\ell}^{\rm pole}(t,0)= {m_{g}\over 2k^{2}}
\,c(t),\end{displaymath}
where
\begin{equation}
c(t)={e^{-im_{g}t}\over 1-e^{-\beta m_{g}}}+{e^{im_{g}t}\over
e^{\beta m_{g}}-1}.\label{3B3}
\end{equation}
Applying the
lemma gives the asymptotic behavior
\begin{equation}
^{*}{\cal D}^{00\,\rm pole}(t,r)\to {-im_{g}\over 8\pi r}
\,c(t)+{\rm exp\; small}.\label{3B4}\end{equation}

\subsubsection{Cut contribution}

For the cut, one needs the small momentum behavior of Eq. (\ref{37}):
\begin{displaymath}
k\to 0:\hskip0.5cm F_{\ell}^{\rm cut}(t,k)\to
T\int_{-k}^{k}\!dk_{0}
\,{\beta_{\ell}(k_{0},k)\over k_{0}}[1+{\cal O}(k_{0}^{2})].
\end{displaymath}
The value of this integral is given by the sum rule in Eq. (\ref{34}).
At small momentum, $Z_{\ell}/\omega_{\ell}\to 1/(2k^{2})$, which makes the
right hand
side finite at $k=0$.    Using the low momentum behavior
of the longitudinal quasiparticle energy,
$\omega_{\ell}^{2}\to m_{g}^{2}+3k^{2}/5+\dots$, in the residue $Z_{\ell}$
gives
for the low momentum limit:
\begin{displaymath}
k\to 0:\hskip0.5cm F_{\ell}^{\rm cut}(t,k)\to {4\over 15}{T\over
m_{g}^{2}}+{\cal O}(k^{2}).
\end{displaymath}
Since this is finite, by the lemma displayed in Eq. (\ref{3B2})
$^{*}{\cal D}^{00\,\rm cut}(t,r)$ does not behave as
$1/r$ in the deep space-like region but instead  falls exponentially with $r$.
The only part of  $^{*}{\cal D}^{00}(t,r)$ that does not fall
exponentially  is the pole contribution already given in  Eq. (\ref{3B4}).
In Appendix \ref{sec:Euclidean} the same result is obtained using the
Euclidean time formulation.

\section{Transverse propagator:
 $^{*}{\cal D}^{\lowercase{ij}}(\lowercase{x})$}

The analysis of  the asymptotic  behavior of the
transverse components of  the HTL propagator  parallels
that of Sec. IV. In the Coulomb gauge the transverse components of the
momentum-space   Wightman function are
\begin{equation}
^{*}D_{>}^{ij}(k_{0},\vec{k})=-i
{\rho_{t}(k_{0},k)\over 1-e^{-\beta k_{0}}}
\,(\delta^{ij}-\hat{k}^{i}\hat{k}^{j}).\label{41}
\end{equation}
Because the spectral integral for  the momentum-space
propagator in Eq. (\ref{2prop}) has poles in $k_{0}$ at the quasiparticle
energies
$\pm\omega_{t}(k)$ and a  branch cut from $k_{0}=-k$ to $k_{0}=k$,
the transverse spectral function  has the structure
\begin{eqnarray}
{1\over 2\pi}\,\rho_{t}(k_{0},k)=&& Z_{t}\big[\delta(k_{0}-\omega_{t})
-\delta(k_{0}+\omega_{t})\big]\nonumber\\
+&&\theta(k^{2}-k_{0}^{2})\,\beta_{t}(k_{0}, k).\nonumber
\end{eqnarray}
The residue function is \cite{P1}
\begin{displaymath}
Z_{t}={\omega_{t}(\omega_{t}^{2}-k^{2})\over
3m_{g}^{2}\omega_{t}^{2}-(\omega_{t}^{2}-k^{2})^{2}}.
\end{displaymath}
The explicit form of the spectral function for the
cut is \cite{P1}
\begin{equation}
\beta_{t}(k_{0},k)={4\over
3m_{g}^{2}}{x(1-x^{2})\over N_{t}(x,k)},\label{43}
\end{equation}
where $x=k_{0}/k$ and the denominator is
\begin{eqnarray}
N_{t}(x,k)\!=&&\!\Big[{4k^{2}\over 3m_{g}^{2}}(1\!-\!x^{2})\!+\!
2x^{2}\!+\!x(1\!-\!x^{2})\ln\!\Big\{\!{1+x\over
1-x}\!\Big\}\!\Big]^{2}\nonumber\\
&&\;+\Big[\pi
x(1-x^{2})\Big]^{2}.\label{44}
\end{eqnarray}
The spectral function for the cut satisfies various sum rules \cite{sum,MLB}.
The one that will be useful here is
\begin{equation}
\int_{-k}^{k}dk_{0}\,{\beta_{t}(k_{0},k)\over k_{0}}
={1\over k^{2}}-{2Z_{t}\over \omega_{t}}.\label{42}
\end{equation}

The  Fourier transform of $^{*}D^{ij}_{>}(k_{o},\vec{k})$ will be written
\begin{eqnarray}
^{*}{\cal D}_{>}^{ij}(t,\vec{r})=&&-i\int{d^{3}k\over (2\pi)^{3}}
\big(\delta^{ij}-\hat{k}^{i}\hat{k}^{j}\big)e^{i\vec{k}\cdot\vec{r}}F_{t}(t,k),
\label{45}\\
F_{t}(t,k)=&&\int_{-\infty}^{\infty}{dk_{0}\over 2\pi}\;{e^{-ik_{0}t}\over
1-e^{-\beta k_{0}}}
\;\rho_{t}(k_{0},k).\nonumber\end{eqnarray}
Here  $F_{t}$ denotes the sum of pole and cut contributions,
$F_{t}=F^{\rm pole}_{t}+F^{\rm cut}_{t}$, where
\begin{equation}
F^{\rm pole}_{t}(t,k)=Z_{t}\bigg[{e^{-i\omega_{t}t}\over
1-e^{-\beta \omega_{t}}}+{e^{i\omega_{t}t}\over
e^{\beta \omega_{t}}-1}\bigg],\label{46}
\end{equation}
\begin{equation}
F^{\rm cut}_{t}(t,k)=\int_{-k}^{k}dk_{0}\;{e^{-ik_{0}t}\over 1-e^{-\beta
k_{0}}}
\;\beta_{t}(k_{0},k).\label{47}
\end{equation}
As in the previous section, it will be important  to know the behavior of
these two functions at small momentum.
As $k\to 0$ the quasiparticle energy   $\omega_{t}\to m_{g}$ and therefore the
residue
$Z_{t}\to 1/2m_{g}$. This implies
\begin{equation}
k\to 0:\hskip0.5cm F_{t}^{\rm pole}(t,0)= {1\over 2m_{g}}\,c(t),\label{48}
\end{equation}
with $c(t)$ the same function as in Eq. (\ref{3B3}).
 For the cut contribution, the sum rule in Eq.
(\ref{42}) and the fact that $Z_{t}\to 1/2m_{g}$ implies that
\begin{equation}
k\to 0:\hskip0.5cm F^{\,\rm
cut}(t,k)\to {T\over k^{2}}-{T\over m_{g}^{2}}+{\cal
O}(k^{2}).\label{49}\end{equation}
 As $k\to 0$ the right hand side diverges as $1/k^{2}$, which will make the
subsequent analysis more difficult.

\paragraph*{Angular integrations and tensor structure:}
It is convenient to first deal with the tensor structure by focusing on the
angular
integrations. The most general rotationally covariant tensor is
\begin{eqnarray}
^{*}{\cal
D}_{>}^{ij}(t,\vec{r})=\big(-\delta^{ij}+3\hat{r}^{i}\hat{r}^{j}\big)&&\,G(t,r)\
nonumber\\
+\big(\delta^{ij}-\hat{r}^{i}\hat{r}^{j}\big)\, &&H(t,r).\label{410}
\end{eqnarray}
By contracting Eq. (\ref{45}) with $\hat{r}_{i}\hat{r}_{j}$ the angular
integration  integration over the solid angle of $\hat{k}$ can be performed
with the result
\begin{eqnarray}
G(t,r)=&&{1\over 2}\hat{r}_{i}\hat{r}_{j}\,^{*}{\cal D}_{>}^{ij}(t,\vec{r})
\nonumber\\
=&&{i\over 2\pi^{2} r}\int_{0}^{\infty}\!\! dk
\;{\partial\over\partial r}
\bigg({\sin kr\over kr}\bigg)\,F_{t}(t,k).\label{411}
\end{eqnarray}
The contraction of Eq. (\ref{45}) with $\delta_{ij}$ leads to
\begin{eqnarray}
H(t,r)=&&{1\over 2}\delta_{ij}\,^{*}{\cal D}_{>}^{ij}(t,\vec{r})\nonumber\\
=&&{-i\over 2\pi^{2} r}\!\int_{0}^{\infty}\!dk\;k\sin(kr) \,F_{t}(t,k).
\label{412}
\end{eqnarray}
The integrands of Eqs. (\ref{411}) and (\ref{412})  both behave at
small $k$ like $k^{2}F_{t}(t,k)$. The previous observations imply that as $k\to
0$,
$k^{2}F^{\rm pole}_{t}(t,k)\to 0$
 and  $k^{2}F^{\rm cut}_{t}(t,k)\to T$. Consequently the integrations for $G$
and $H$ are convergent at small $k$. By using the relation
\begin{displaymath}
\Big[{\partial^{2}\over\partial r^{2}}+{2\over
r}{\partial\over\partial r}+k^{2}\Big]\!\Big({\sin kr\over kr}\Big)=0,
\end{displaymath}
it is easy to show that
the two coefficient functions are related by
\begin{equation}
{\partial\over\partial r}\Big(r^{3}G(t,r)\Big)=r^{2}H(t,r).\label{413}
\end{equation}
It will often be easier to compute the integral for $H(t,r)$ in Eq. (\ref{412})
and then use this relation to compute $G(t,r)$. Of course it will not
determine any part of $G$ that behaves like $1/r^{3}$.

The asymptotic behaviors of $G$ and $H$ will be computed in this section,
but the results will not be substituted into Eq. (\ref{410}) for examination.
That examination and discussion will be deferred to Sec. VI and the impatient
reader interested in the answers rather than the computations is encouraged
to  turn directly to Sec. VI.

\subsection{Asymptotic time-like behavior of
 $\;^{*}{\cal D}^{\lowercase{ij}}(\lowercase{x})$}

The  deep time-like region denotes the limit in which  $t\to \infty$,
$r\to\infty$ at a fixed ratio satisfying $t/r> 1$.

\subsubsection{Pole contribution} The asymptotic behavior of the pole
contribution to the transverse propagator is  controlled by a point of
stationary
phase. From Eqs. (\ref{46}) and (\ref{412}) the contribution of the
quasiparticle pole is
 \begin{eqnarray}
H^{\rm pole}(t,r)=&&{-i\over 2\pi^{2}r}\int_{0}^{\infty}\!dk\;
k\,\sin kr\nonumber\\
&&\times Z_{t}\bigg[{e^{-i\omega_{t}t}\over
1-e^{-\beta \omega_{t}}}+{e^{i\omega_{t}t}\over
e^{\beta \omega_{t}}-1}\bigg].\nonumber
\end{eqnarray}
 One can
extend the integration to negative $k$ so that
 \begin{eqnarray}
H^{\rm pole}(t,r)=&&{-1\over 4\pi^{2}r}\int_{-\infty}^{\infty}\!dk\,
k Z_{t}\bigg[{e^{ikr-i\omega_{t}t}\over
1-e^{-\beta \omega_{t}}}-{e^{-ikr+i\omega_{t}t}\over
e^{\beta \omega_{t}}-1}\bigg].\nonumber
\end{eqnarray}
As before, the method of stationary phase
\cite{cop,din} will be used to compute the asymptotic value of this
integral. Define
\begin{equation}
\phi(k)=kr-\omega_{t}(k) t.\label{tphase}\end{equation}
At large values of $r$ and $t$ the functions $e^{\pm i\phi}$ oscillate
rapidly except at special values of $k$.  Let $\overline{k}$ be the value of
the wave vector at which the phase is stationary.
This requires that the group velocity of the transverse quasiparticle satisfy
\begin{equation}
\bigg[{d\omega_{t}\over dk}
\bigg]_{\overline{k}}={r\over t}< 1.
\end{equation}
The group velocity of the transverse dispersion relation is
zero at
$k=0$ and grows monotonically with $k$, approaching 1 as $k\to\infty$
\cite{MLB}. Therefore there is  always a real
$\overline{k}$  which satisfies the stationary phase condition.  Near the
stationary point, the phase is
\begin{displaymath}
\phi(k)=\phi(\overline{k})-{1\over
2}(k-\overline{k})^{2}\,t\,\bigg[{d^{2}\omega_{t}\over
dk^{2}}\bigg]_{\overline{k}}+\dots.
\end{displaymath}
The acceleration $d^{2}\omega_{t}/dk^{2}$ is always positive.
The $k$ contour must be deformed into the complex plane so that it passes
through the real value of $\overline{k}$ on a path of
steepest descent. For the first integrand,
$\exp(+i\phi)$, this requires $k=\overline{k}+e^{-i\pi/4}s$
where $s$ is real.
For the second integrand, $\exp(-i\phi)$, the correct path is
$k=\overline{k}+e^{i\pi/4}s$. Performing the two Gaussian integrations gives
the asymptotic behavior:
\begin{eqnarray}
H^{\rm pole}(t,r)\!\to\! &&{-\overline{k}\,\overline{Z}_{t}\over
r\sqrt{t\,\overline{\omega}_{t}''}}\label{4A3}\\
&&\times {1\over (2\pi)^{3/2}}\bigg[{e^{i\overline{\phi}-i\pi/4}\over
1-e^{-\beta\overline{\omega}_{t}}}-{e^{-i\overline{\phi}+i\pi/4}
\over e^{\beta\overline{\omega}_{t}}-1}\bigg]\nonumber\end{eqnarray}
Since the ratio $r/t$ is fixed, this falls like $t^{-3/2}$.

The tensor structure of the transverse propagator in Eq. (\ref{410})
depends on both
$H$ and $G$. Using Eq. (\ref{413}) gives for the asymptotic behavior of
$G^{\rm pole}$:
\begin{eqnarray}
G^{\rm pole}(t,r)\!\to\! &&{i\,\overline{Z}_{t}\over
r^{2}\sqrt{t\,\overline{\omega}_{t}''}}\label{4A4}\\
&&\times {1\over (2\pi)^{3/2}}\bigg[{e^{i\overline{\phi}-i\pi/4}\over
1-e^{-\beta\overline{\omega}_{t}}}+{e^{-i\overline{\phi}+i\pi/4}
\over e^{\beta\overline{\omega}_{t}}-1}\bigg]\nonumber\end{eqnarray}
To confirm that this is the dominant term in $G^{\rm pole}$ it is
necessary to recognize that
the most rapid dependence on $r$ comes from  the phase,
$\overline{\phi}=\overline{k}r-\omega_{t}(\overline{k})t$.
The $r$ derivative of $G^{\rm pole}$ comes from three sources:
from the  dependence of the phase $\overline{\phi}$ on r,
from the $1/r^{2}$
power dependence,   and from
the implicit dependence of the stationary point $\overline{k}$ on $r$:
\begin{displaymath}
{\partial G^{\rm pole}\over \partial r}={d\overline{\phi}\over d
r}\,{\partial G^{\rm pole}
\over\partial\overline{\phi}}-{2\over r} G^{\rm
pole}+{d\overline{k}\over dr}{\partial G^{\rm pole}
\over\partial \overline{k}}.
\end{displaymath}
Since
$d\overline{\phi}/dr=\overline{k}$
the first term is  larger than the second at large $r$. The first term is
also larger than the third term because
 dependence of $\overline{k}$ on  $r$ is very slow:
$d\overline{k}/dr=1/(t\overline{\omega}_{t}'')$, which is of order $1/r$.
This confirms Eq. (\ref{4A4}).

\subsubsection{Cut contribution}
To analyze the cut contribution of the transverse propagator in the deep
time-like
region it is convenient to begin with the
double integral representation
\begin{eqnarray}
H^{\rm cut}(t,r)\!=&&{-i\over
r}\int_{-\infty}^{\infty}\!{dk\over (2\pi)^{2}}\,k\sin
kr\nonumber\\
&&\times\int_{-k}^{k}\!dk_{0}
\;{e^{-ik_{0}t}\over 1-e^{-\beta k_{0}}}\,\beta_{t}(k_{0},k),
\nonumber\end{eqnarray}
which follows from Eqs. (\ref{47}) and (\ref{412}).
As previously noted, the integral over $k_{0}$ approaches $T/k^{2}$ at
small $k$
and therefore the integration is  convergent for small $k$.
Changing variables from   $k_{0}$  to
the phase velocity $x=k_{0}/k$
and  replacing  $i\sin kr$  with $\exp(ikr)$ gives
\begin{eqnarray}
H^{\rm cut}(t,r)\!=&&{-1\over
r(2\pi)^{2}}\int_{-\infty}^{\infty}\!dk\int_{-1}^{1}\!dx
\;e^{ik(r-xt)}\,I_{t}(x,k)\nonumber\\
I_{t}(x,k)=&&{k^{2}\beta_{t}(xk,k)\over 1-e^{-\beta xk}}
.\label{4AIt}\end{eqnarray}
This double integral is quite similar
to the integral which arose for the longitudinal propagator in Eq.
(\ref{3A4}). The stationary phase point occurs at
$\overline{x}=r/t$,
$\overline{k}=0$. The contribution of this stationary phase point will be of
order $1/r^{3}$, which was the final result  for the longitudinal propagator
in Eq. (\ref{3Alast}).

It is rather surprising that the stationary phase point does not give the
largest term in the asymptotic expansion. The subtlety comes ultimately
from the
fact that there is no magnetic screening.
There is a hand-waving way to get the correct leading behavior.
Reference to Eq. (\ref{49}) suggests making the split
\begin{displaymath}
F_{t}^{\rm cut}(t,k)={T\over k^{2}}+\delta F_{t}^{\rm cut}(t,k).
\end{displaymath}
This gives an elementary integration for the leading term:
\begin{eqnarray}
H^{\rm cut}(t,r)=
&&{-i\over 2\pi^{2} r}\!\int_{0}^{\infty}\!dk\;k\sin(kr) \bigg[{T\over
k^{2}}+\delta F_{t}^{\rm cut}(t,k)\bigg]\nonumber\\
=&&{-iT\over 4\pi r}-{i\over 2\pi^{2} r}\!\int_{0}^{\infty}\!dk\;k\sin(kr)
\,\delta F_{t}^{\rm cut}(t,k).
\nonumber\end{eqnarray}
It will turn out that $-iT/4\pi r$ actually is the correct leading term.
However
this simple calculation does not really prove that  because one cannot show
that the contribution of the integral over $\delta F_{t}^{\rm cut}$ is smaller
than $T/r$. The problem is that $\delta F_{t}^{\rm cut}$ is too awkward to
deduce the asymptotic behavior.

A more convincing analysis
 is to check qualitatively that there is a
region of integration whose contribution is larger than the point of
stationary phase. Thus we examine the integrand $I_{t}$ of the
double integral Eq. (\ref{4AIt}) in the region where $x\ll 1$ and  $|k|\ll
m_{g}^{2}$:
\begin{displaymath}
I_{t}(x,k)\bigg|_{\rm small\; x, small \;k}\approx {4T\over
3m_{g}^{2}}\;{k\over
\big[4k^{2}/3m_{g}^{2}\big]^{2}+\big[\pi x\big]^{2}}.
\end{displaymath}
This part of the integrand is largest when $k^{2}/m_{g}^{2}$ is similar in
magnitude to $x$. The large value of $I_{t}$ can be suppressed by
oscillations in the phase factor
$\exp[ik(r-xt)]$. However, at very small  $k\ll 1/r$ these oscillations are
negligible.  Thus the dominant region is $k\sim m_{g}\sqrt{x}\ll 1/r$. The
approximate value of the integrand here is very large:
$I_{t}(x,k)\sim Tm_{g}^{2}r^{3}$
The effect of this  small region on $H^{\rm cut}$ can be crudely
estimated as
\begin{equation}
H^{\rm cut}(t,r)\sim {1\over
r}\int_{0}^{r^{-1}}\!\!\!dk\int_{0}^{(m_{g}r)^{-2}}
\!\!\!dx\;\big[Tm_{g}^{2}r^{3}\big]\sim {T\over r}.
\end{equation}
At sufficiently large $r$ this dominates the stationary phase contribution,
$T/(m_{g}^{2}r^{3})$. Note also that the $T/r$ contribution is independent
of $m_{g}$ and thus independent of the coupling.

To perform a precise  calculation of the leading and sub-leading  behavior of
$H^{\rm cut}$, the essential step is to separate out the
dominant contribution to the integrand by defining
\begin{eqnarray}
I^{(1)}(x,k)=&&{4T\over 3m_{g}^{2}}\;{k\over
\big[4k^{2}/3m_{g}^{2}\big]^{2}+\big[\pi x\big]^{2}}\label{4Ai1}\\
I^{(2)}(x,k)=&&I_{t}(x,k)-I^{(1)}(x,k).\label{4Ai2}
\end{eqnarray}
Obviously $I^{(1)}$ contains the most important behavior.
With this split,  $H^{\rm cut}=H^{(1)}+H^{(2)}$ where for
$n=1$ or $2$ the  integrals are
given by
\begin{displaymath}
H^{(n)}(t,r)\!={-1\over
r(2\pi)^{2}}\int_{-\infty}^{\infty}\!dk\int_{-1}^{1}\!dx
\;e^{ik(r-xt)}\,I^{(n)}(x,k).
\end{displaymath}

(1) The dominant contribution will come from $H^{(1)}$.  This is computed
by direct integration in Appendix \ref{sec:cut}. That calculation gives
the asymptotic behavior
\begin{equation}
H^{(1)}(t,r)\to {-iT\over 4\pi r}-{4iT\over
3\pi^{3}m_{g}^{2}r^{3}\overline{x}},
\label{4H1}\end{equation}
with power-law corrections.
Using this in Eq. (\ref{413}) gives the asymptotic behavior of $G^{(1)}$ as
\begin{equation}
G^{(1)}(t,r)\to {-iT\over 8\pi r}+{4iT\over 3\pi^{3}m_{g}^{2}r^{3}\overline{x}}
.\label{4G1}
\end{equation}

(2) The  behavior of $H^{(2)}$ in the deep time-like region
can be obtained from the same type of stationary-phase method used for
$^{*}{\cal D}_{>}^{00
\,\rm cut}(t,r)$ in Sec. IV A.  The change of variables in Eq. (\ref{uv})
gives
\begin{displaymath}
H^{(2)}(t,r)\!={-2\over
rt(2\pi)^{2}}\!\int_{0}^{u_{\rm max}}\!udu\int_{0}^{2\pi}\!d\theta
\;e^{-u^{2}}\,I^{(2)}(x,k).\end{displaymath}
At the stationary point, $\overline{x}=r/t$ and $\overline{k}=0$,
the integrand $I^{(2)}(\overline{x},0)$ vanishes. The leading term is obtained
using  Eq. (\ref{3A6}):
\begin{displaymath}
H^{(2)}(t,r)\!\to{-2\over
rt(2\pi)^{2}}
\Big({-i\pi\over t}\Big)
\Big[{\partial^{2} I^{(2)}\over\partial x\partial k}\Big]_{
\overline{x},0}.\end{displaymath}
Working this out explicitly gives
\begin{equation}
H^{(2)}(t,r)\!\to{i2T\over
rt^{2}3\pi m_{g}^{2}}
\,{d\over d \overline{x}}\bigg[{1-\overline{x}^{2}
\over N_{t}(\overline{x},0)}-{1\over
\pi^{2}\overline{x}^{2}}\bigg].\label{4H2}\end{equation}
Because of the subtracted term this is finite even at $\overline{x}=0$.
From Eq. (\ref{413}) the other coefficient $G^{(2)}$ must have the
asymptotic behavior
\begin{equation}
G^{(2)}(t,r)\to {2iT\over 3\pi m_{g}^{2}r^{3}}f(\overline{x}),
\label{4G2}\end{equation}
where $f(\overline{x})$ satisfies
\begin{displaymath}
{d f(\overline{x})\over d\overline{x}}=
\overline{x}\,{d\over d \overline{x}}\bigg[{1-\overline{x}^{2}
\over N_{t}(\overline{x},0)}-{1\over
\pi^{2}\overline{x}^{2}}\bigg].
\end{displaymath}
Integrating this is not automatic because one does not have
an initial condition.
As $\overline{x}\to 0$ the right hand side is of order $\overline{x}^{2}$.
Thus $f(\overline{x}) \to f(0)+{\cal O}(\overline{x}^{3})$ as
$\overline{x}\to 0$. It is not obvious that $f(0)$ should vanish. (This
would make $G^{(2)}$ finite at $r=0$, but there is no reason to demand this
since
$G^{(1)}$ is not finite at $r=0$.)
Appendix \ref{sec:cut} performs a stationary phase calculation of $G^{(2)}$
with the result
\begin{displaymath}
f(\overline{x})=
\int_{0}^{\overline{x}}\! dy\; y\,{d\over dy}\bigg[
{1-y^{2}\over N_{t}(y,0)}-{1\over (\pi y)^{2}}\bigg].
\end{displaymath}
In the calculation of Appendix \ref{sec:cut} the integration
variable over $y$ is a remnant of the angular integration. Specifically,
$y=\overline{x}\cos\theta$ where
$\hat{k}\cdot\hat{r}=\cos\theta$.
 Since the quantity in square brackets
is finite at $y=0$, one can integrate by parts to obtain
\begin{displaymath}
f(\overline{x})\!=\!
{\overline{x}(1\!-\!\overline{x}^{2})\over N_{t}(\overline{x},0)}
\!-\!{1\over \pi^{2}\overline{x}}
\!-\!\int_{0}^{\overline{x}}\! dy\bigg[
{1\!-\!y^{2}\over N_{t}(y,0)}\!-\!{1\over (\pi y)^{2}}\bigg].
\end{displaymath}
In this form the entire integral of the last term $1/(\pi y)^{2}$ appears
important. Because the expression in square brackets is finite at $y=0$, one
can replace the lower limit of zero by an infinitesimal value $y_{0}$ and
take the limit $y_{0}\to 0$ after performing the integration.
Doing this gives
\begin{displaymath}
f(\overline{x})=
{\overline{x}(1-\overline{x}^{2})\over N_{t}(\overline{x},0)}
-{2\over \pi^{2}\overline{x}}
-\lim_{y_{0}\to 0}\bigg[\int_{y_{0}}^{\overline{x}}\! dy\;
{1-y^{2}\over N_{t}(y,0)}-{1\over \pi^{2}y_{0}}\bigg].
\end{displaymath}
In this form the divergence from the lower limit of the integration is
 canceled by the term $-1/(\pi^{2}y_{0})$.

\paragraph*{Total cut contribution.} In the last form for $f(\overline{x})$ the
term $-2/(\pi^{2}\overline{x})$ when substituted into Eq. (\ref{4G2}) exactly
cancels the order $1/r^{3}$ term in Eq. (\ref{4G1}).
 Consequently the sum of $G^{(1)}$ and $G^{(2)}$ is
\begin{eqnarray}
&&G^{\rm cut}(t,r)\to {-iT\over 8\pi r}+{2iT\over 3\pi
m_{g}^{2}r^{3}}\,g(\overline{x})\\
&&g(\overline{x})= {\overline{x}(1-\overline{x}^{2})\over
N_{t}(\overline{x},0)}
-\lim_{y_{0}\to 0}\bigg[\int_{y_{0}}^{\overline{x}}\!dy{1-y^{2}
\over N_{t}(y,0)}-{1\over\pi^{2}y_{0}}\bigg].\nonumber
\end{eqnarray}
The surviving terms of order $1/r^{3}$
come entirely from the stationary phase point.
Similarly when $H^{(1)}$ and $H^{(2)}$  are added together the order $1/r^{3}$
term from Eq. (\ref{4H1}) is canceled by the last term in Eq. (\ref{4H2})
to give
\begin{equation}
H^{\rm cut}(t,r)\to {-iT\over 4\pi r}+
{i2T\over rt^{2}3\pi m_{g}^{2}}
{\partial\over\partial \overline{x}}\bigg[{1-\overline{x}^{2}
\over N_{t}(\overline{x},0)}\bigg].\label{4Hcut}
\end{equation}
The corrections to $1/r$
come from the stationary phase point.

\paragraph*{Inclusion of a magnetic mass.} It is widely believed
that non-abelian gauge theories at high temperature generate a magnetic mass of
order
$g^{2}T$ defined by the limit \cite{L,GPY}
\begin{displaymath}
\mu^{2}=\lim_{k\to 0}\Pi_{t}(0,k).\end{displaymath}
It is not known  how a magnetic mass would change the transverse spectral
function at nonzero values of $k_{0}$ and $k$.   The simplest possibility
is that
the occurrence of $k^{2}$ in the transverse spectral function is replaced by
the sum $k^{2}+\mu^{2}$. Thus the denominator of the new spectral function,
denoted with a tilde, is
\begin{eqnarray}
\tilde{N}_{t}(x,k)\!=&&\!\Big[{4(k^{2}\!+\!\mu^{2})\over
3m_{g}^{2}}(1\!-\!x^{2})\!+\!
2x^{2}\!+\!x(1\!-\!x^{2})\ln\!\Big\{\!{1\!+\!x\over
1\!-\!x}\!\Big\}\!\Big]^{2}\nonumber\\  &&\;+\Big[\pi
x(1-x^{2})\Big]^{2}.\label{428}
\end{eqnarray}
The  function $H^{\rm cut}$ becomes
\begin{displaymath}
\tilde{H}^{\rm cut}(t,r)\!={-1\over
r(2\pi)^{2}}\int_{-\infty}^{\infty}\!dk\int_{-1}^{1}\!dx
\;e^{ik(r-xt)}\,\tilde{I}_{t}(x,k),\end{displaymath}
where the magnetic mass now occurs in the denominator of the spectral function:
\begin{displaymath}
\tilde{I}_{t}(x,k)={4\over 3m_{g}^{2}}{k^{2}x(1-x^{2})\over [1-e^{-\beta kx}]
\tilde{N}_{t}(x,k)}.
\end{displaymath}
Because of the magnetic mass $\tilde{I}_{t}(x,k)$ does not diverge at
 $x\sim\sqrt{k/m_{g}}\to 0$.

The asymptotic behavior of $\tilde{H}^{\rm cut}$ in the deep time-like region
can be obtained from the stationary-phase method used in Sec. III A.  The
change of variables in Eq. (\ref{uv}) gives
\begin{displaymath}
\tilde{H}^{\rm cut}(t,r)\!={-2\over
rt(2\pi)^{2}}\int_{0}^{u_{\rm max}}\!udu\int_{0}^{2\pi}\!d\theta
\;e^{-u^{2}}\,\tilde{I}_{t}(x,k).\end{displaymath}
At the stationary-phase point,  $\tilde{I}_{t}(\overline{x},0)$ vanishes
and so according to Eq. (\ref{3A6}) the leading term is
\begin{displaymath}
\tilde{H}^{\rm cut}(t,r)\!\to{-2\over
rt(2\pi)^{2}}
\Big({-i\pi\over t}\Big)
\Big[{\partial^{2} \tilde{I}_{t}\over\partial x\partial k}\Big]_{
\overline{x},0}.\end{displaymath}
Working this out explicitly gives
\begin{equation}
\tilde{H}^{\rm cut}(t,r)\!\to{i2T\over
rt^{2}3\pi m_{g}^{2}}
{\partial\over\partial \overline{x}}\bigg[{1-\overline{x}^{2}
\over \tilde{N}_{t}(\overline{x},0)}\bigg].\label{429}\end{equation}
The asymptotic behavior of $\tilde{G}^{\rm cut}$ is
\begin{eqnarray}
\tilde{G}^{\rm cut}(t,r)\!\to\!
{2iT\over 3\pi
m_{g}^{2}r^{3}}\! \bigg[{\overline{x}(1\!-\!\overline{x}^{2})\over
\tilde{N}_{t}(\overline{x},0)}
\!-\!\int_{0}^{\overline{x}}\!dy\,{1\!-\!y^{2}
\over \tilde{N}_{t}(y,0)}\bigg].\label{430}
\end{eqnarray}
Thus in the presence of a magnetic mass the transverse propagator falls like
$1/t^{3}$ in the deep time-like region.

\subsection{Asymptotic space-like behavior of
 $\;^{*}{\cal D}^{\lowercase{ij}}(\lowercase{x})$}

The limit $t\to\infty$,
$r\to\infty$ with a fixed ratio $t/r< 1$ is the deep space-like region.
As observed in Sec. IV B,   there are  no points of stationary phase
for real values of $k$ and $k_{0}$. Stationary phase points at complex values
of $k$ or $k_{0}$ will produce terms that fall exponentially.
The largest effects are instead endpoint contributions that come from the
region of small $k$ in the integral of Eq. (\ref{411}):
\begin{equation}
G(t,r)=
{i\over 2\pi^{2} r}\int_{0}^{\infty}\!\! dk
\;{\partial\over\partial r}
\bigg({\sin kr\over kr}\bigg)\,F_{t}(t,k).\label{4B1}
\end{equation}

\subsubsection{Pole contribution} Eq. (\ref{48}) shows that at zero momentum,
$F_{t}^{\rm pole}(t,0)=c(t)/2m_{g}$. The $r$ derivative may be taken
outside the integral defining $G^{\rm pole}$ so that
\begin{displaymath}
G^{\rm pole}(t,r)={i\over 2\pi^{2} r}{\partial\over\partial
r}\bigg[{1\over r}\int_{0}^{\infty}\!\! dk
\;{\sin kr\over k} F_{t}(t,k)\bigg].
\end{displaymath}
The integral over $k$ now satisfies the conditions of the lemma
displayed in Eq. (\ref{3B2}) and proven in Appendix \ref{sec:lemma}. Therefore
in the deep space-like limit
\begin{equation}
G^{\rm pole}(t,r)\to {i\over 2\pi^{2} r}{\partial\over\partial
r}\bigg[{1\over r}{\pi\over 2}{c(t)\over 2m_{g}}\bigg]
={-ic(t)\over 4\pi m_{g}r^{3}}.
\end{equation}
The corrections to this are exponentially small. Since this is a pure
$1/r^{3}$ behavior, by Eq. (\ref{413}) the asymptotic behavior of the other
coefficient function is
\begin{equation}
H^{\rm pole}(t,r)\to {\rm exp\; small}.
\end{equation}

\subsubsection{Cut contribution} For the contribution of the cut,
the result in Eq. (\ref{49}) motivates  separating out the dominant behavior
by defining
\begin{displaymath}
F_{t}^{\,\rm cut}(t,r)={T\over k^{2}}+\delta F_{t}^{\,\rm cut}(t,k),
\end{displaymath}
where $\delta F_{t}^{\,\rm cut}(t,k)$ is finite at $k=0$.
Using this in Eq. (\ref{4B1}) gives
\begin{eqnarray}
G^{\rm cut}(t,r)=&&{i\over 2\pi^{2} r}\int_{0}^{\infty}\! dr
{\partial\over\partial r}\bigg({\sin kr\over kr}\bigg)
{T\over k^{2}}\nonumber\\
+&&{i\over 2\pi^{2} r}{\partial\over \partial r}
\int_{0}^{\infty}\! dk\,{\sin kr\over kr}\,\delta F^{\rm cut}_{t}(t,k).
\nonumber
\end{eqnarray}
Both these integrals are convergent at $k=0$ and at $k=\infty$.
The first integral may be done exactly. For the second, the asymptotic behavior
is given by the lemma, Eq. (\ref{3B2}).
The result is
\begin{equation}
G^{\rm cut}(t,r)\to {-iT\over 8\pi r}+{iT\over 4\pi m_{g}^{2} r^{3}}
+{\rm exp\;small}.\label{4BGcut}
\end{equation}
By Eq. (\ref{413}) the other coefficient function has asymptotic behavior
\begin{equation}
H^{\rm cut}(t,r)\to {-iT\over 4\pi r}+{\rm exp.\;small}.
\label{4BHcut}\end{equation}

\paragraph*{Inclusion of a magnetic mass.} The existence of a magnetic mass
$\mu^{2}$ would change  the asymptotic space-like behavior of the
transverse propagator. The contribution of the quasiparticle pole
 does  not change appreciably. However, for the contribution of the
branch cut the sum-rule in Eq. (\ref{42}) changes to
\begin{displaymath}
\int_{-k}^{k}dk_{0}\,{\beta_{t}(k_{0},k)\over k_{0}}={1\over
k^{2}+\Pi_{t}(0,k)}-{2Z_{t}\over
\omega_{t}},\end{displaymath}
as argued by Pisarski \cite{sum}.
In the limit of small $k$ this gives
\begin{displaymath}
k\to 0:\;\int_{-k}^{k}dk_{0}\,{\beta_{t}
(k_{0},k)\over k_{0}}\to {1\over \mu^{2}}-{1\over
m_{g}^{2}}+{\cal O}(k^{2}).
\end{displaymath}
Therefore in Eq. (\ref{4BGcut}) the term of order $1/r$ is absent and the
coefficient of the $1/r^{3}$ changes so that
\begin{eqnarray}
\tilde{G}^{\rm cut}(t,r)\to {iT\over 4\pi r^{3}}\bigg[
{1\over m_{g}^{2}}-{1\over\mu^{2}}\bigg] +{\rm exp.\; small}.\label{436}
\end{eqnarray}
By Eq. (\ref{413})
\begin{equation}
\tilde{H}^{\rm cut}(t,r)\to  {\rm exp\; small}.\label{437}
\end{equation}
Among the exponentially small terms is the familiar $\exp(-\mu r)$.
However, the dominant behavior is the $1/r^{3}$ term derived here.
In Appendix \ref{sec:Euclidean} the same results are obtained using the
imaginary time approach.

\section{Summary and Discussion}

This section reviews the results obtained in the previous calculations and
compares
them with free propagators at finite temperature. The summary is
self-contained and may be read independently of the previous sections.
It is organized differently than the text.
First comes the asymptotic time-like behavior of thermal propagators; then
the asymptotic space-like behaviors. The propagators are always in the Coulomb
gauge.

It is perhaps worth repeating that ghost propagators do not contain
hard-thermal-loops and therefore are not resummed \cite{BP1}. In the HTL
approximation the ghost propagators are just the free thermal propagators.
Their asymptotic behavior may be found in \cite{AW}.

\begin{table}
\caption{Leading asymptotic behavior of gauge boson propagators at finite
temperature in the Coulomb gauge.}
\begin{tabular}{lcc}
 Type &  Time-like  & Space-like\\
\tableline
Free massless & ${\rm exp.\;small}$   &  $1/r$  \\
HTL $^{*}{\cal D}^{00}$  pole & $1/t^{3/2}$ &  $1/r$\\
HTL $^{*}{\cal D}^{00}$ cut & $1/t^{3}$ & exp small \\
HTL $^{*}{\cal D}^{ij}$ pole & $ 1/t^{3/2}$ & $1/r^{3}$\\
HTL $^{*}{\cal D}^{ij}$ cut
 &  $1/r$ \tablenote {This   changes to
$1/t^{3}$ if there is a magnetic mass.}
 & $1/r$ \tablenote {This   changes to
$1/r^{3}$ if there is a magnetic mass.}
\end{tabular}\label{two}
\end{table}

\subsection{Asymptotic time-like behavior of gauge boson propagators}

In the deep time-like region $t$ and $r$ are both large with a fixed ratio
$t/r>1$. For the free propagator $t$ and $r$ must be large compared to
$1/2\pi T$. For the HTL propagator, they need to be large compared to
$1/m_{g}$.  The leading term in the asymptotic time-like behavior is
displayed in
the first column of Table I.

\paragraph*{1. Free, massless gauge boson.} In Coulomb gauge the scalar
potential
$A^{0}(x)$ is instantaneous and consequently
${\cal D}^{00}(t,r)=\delta(t)/4\pi r$.
The exact space-time dependence of
${\cal D}^{ij}(t,\vec{r})$ can be found
in \cite{AW}.  In the deep time-like region
$t>r\gg 1/(2\pi T)$ the results are
\begin{mathletters}\begin{eqnarray}
{\cal D}^{00}(t,r)=&& 0\\
{\cal D}^{ij}(t,\vec{r})\to&& {\rm exp.\; small.}.
\end{eqnarray}\label{61}\end{mathletters}
Exponential fall-off in the time-like direction is characteristic of free
massless
bosons and was found in the Feynman gauge example in Sec. II. The same
behavior occurs for spin 0, for spin 1 in any gauge, and for spin 2 gravitons
\cite{AW}. The exponential fall-off can be understood in terms of the
calculations
performed in Secs. IV A and V A.  The  power-law behavior that occurs in
the HTL
propagators comes from a point of stationary phase. The stationary phase
contribution  always has exponentially small corrections, which have been
consistently omitted.  However, for free, massless bosons  since
$k_{0}=k$, there can be no value of $k$ for which the phase factor
$\exp(ik(r-t))$ is stationary. The entire answer consists of the exponentially
small corrections.

\paragraph*{2. HTL longitudinal propagator.} From Sec. IV A the asymptotic
time-like behavior of the longitudinal gauge propagator for $t\gg 1/m_{g}$
is
\begin{mathletters}\begin{eqnarray}
^{*}{\cal D}^{00\,\rm pole}(t,r)\to&& {a_{\ell}^{\rm pole}(t,r)\over t^{3/2}}\\
^{*}{\cal D}^{00\,\rm cut}(t,r)\to&&-i {a_{\ell}^{\rm cut}(\overline{x})\over
t^{3}},\end{eqnarray}\label{62}\end{mathletters}
where the coefficient functions are
\begin{eqnarray}
a_{\ell}^{\rm pole}(t,r)=&& {-\overline{k}\;\overline{Z}_{\ell}
\over \overline{x}\sqrt{(2\pi)^{3}\,\overline{\omega}_{\ell}''}}
\bigg[{e^{i\overline{\phi}-i\pi/4}\over
1-e^{-\beta\overline{\omega}_{\ell}}} -{e^{-i\overline{\phi}+i\pi/4}\over
e^{\beta\overline{\omega}_{\ell}}-1}\bigg],\nonumber\\
a_{\ell}^{\rm cut}(\overline{x})=&&{-T\over 3\pi m_{g}^{2}\overline{x}}
{d\over d\overline{x}}\bigg[{1\over
N_{\ell}(\overline{x},0)}\bigg].\nonumber\end{eqnarray}
The function $a_{\ell}^{\rm cut}$ depends only on the  variable
$\overline{x}=r/t$. The function $a_{\ell}^{\rm pole}$ also depends on the
ratio
$r/t$, since the stationary point is the value momentum $\overline{k}$ at
which the group velocity of the quasiparticle satisfies
$d\omega_{\ell}/dk=r/t$. In addition $a_{\ell}^{\rm pole}$ depends  on $t$
and $r$ through the
  value of the phase at the
stationary point: $\overline{\phi}=\overline{k}r-\omega_{\ell}(\overline{k})t$.

\paragraph*{3. HTL transverse propagator.} From Sec. V A the  transverse
gauge propagator in the deep time-like region has the following
asymptotic behavior:
\begin{mathletters}\begin{eqnarray}
^{*}{\cal D}^{ij\,\rm pole}(t,\vec{r})\to &&{b_{t}^{\rm pole}(t,r)
\over t^{5/2}}
(-\delta^{ij}+3\hat{r}^{i}\hat{r}^{j}) \nonumber\\
+&& {b_{t}^{\prime\rm pole}(t,r)\over t^{3/2}}
(\delta^{ij}-\hat{r}^{i}\hat{r}^{j}) \\
^{*}{\cal D}^{ij\,\rm cut}(t,\vec{r})\to &&{-iT\over 8\pi r}
(\delta^{ij}+\hat{r}^{i}\hat{r}^{j})\nonumber \\
-&&i{b_{t}^{\rm
cut}(\overline{x})\over t^{3}}
(-\delta^{ij}+3\hat{r}^{i}\hat{r}^{j})\nonumber\\ -&&i{b_{t}^{\prime\rm
cut}(\overline{x})\over t^{3}} (\delta^{ij}-\hat{r}^{i}\hat{r}^{j}).
\end{eqnarray}\label{63}\end{mathletters}
for times $t\gg 1/m_{g}$.
The  hard-thermal-loop propagators do not contain a magnetic mass.
Therefore Eqs. (\ref{63}) apply to high temperature QED.
For high temperature QCD the inclusion of a magnetic mass changes the
asymptotic behavior as discussed in the next paragraph.
The coefficients for the pole contribution are
\begin{eqnarray}
b_{t}^{\rm pole}(t,r)\!=\! &&{i\,\overline{Z}_{t}\over
\overline{x}^{2}\sqrt{(2\pi)^{3}\overline{\omega}_{t}''}}
\bigg[{e^{i\overline{\phi}-i\pi/4}\over
1-e^{-\beta\overline{\omega}_{t}}}+{e^{-i\overline{\phi}+i\pi/4}
\over e^{\beta\overline{\omega}_{t}}-1}\bigg],\nonumber\end{eqnarray}
\begin{eqnarray}
b_{t}^{\prime\rm pole}(t,r)\!=\! {-\overline{k}\,\overline{Z}_{t}\over
\overline{x}\sqrt{(2\pi)^{3}\overline{\omega}_{t}''}}
\bigg[{e^{i\overline{\phi}-i\pi/4}\over
1-e^{-\beta\overline{\omega}_{t}}}-{e^{-i\overline{\phi}+i\pi/4}
\over e^{\beta\overline{\omega}_{t}}-1}\bigg].\nonumber\end{eqnarray}
These separately on  $t$ and $r$ through the phase $\overline{\phi}$.
 The coefficients for the cut contributions depend only on the ratio
$\overline{x}=r/t$:
\begin{eqnarray}
b_{t}^{\rm cut}(\overline{x})=&& {2T\over 3\pi m_{g}^{2}\overline{x}^{3}}
\bigg\{
\lim_{y_{0}\to 0}\bigg[\int_{y_{0}}^{\overline{x}}\!dy{1-y^{2}
\over N_{t}(y,0)}-{1\over\pi^{2}y_{0}}\bigg]\nonumber\\
&&\hskip4cm -{\overline{x}(1-\overline{x}^{2})\over
N_{t}(\overline{x},0)}\bigg\},\nonumber\\
b_{t}^{\prime\rm cut}(\overline{x})=&&
{-2T\over 3\pi m_{g}^{2}\overline{x}}\,
{\partial\over\partial \overline{x}}\bigg[{1-\overline{x}^{2}
\over N_{t}(\overline{x},0)}\bigg].\nonumber
\end{eqnarray}
In Sec. V A the calculation of the leading term,
 $T/r$, in  $^{*}{\cal D}^{ij\,\rm cut}(t,\vec{r})$ was lengthy.
However, it has a profound implication for the behavior of electrons.
As shown by Pisarski the electron propagator dressed with one HTL photon has
a damping rate that diverges on the electron mass shell \cite{sum}. In
space-time this comes from the $T/r$. Blaizot and Iancu showed that summing
the most infrared-sensitive contributions to the electron propagator
produces a time dependence of the form $\exp[-\alpha Tt\ln(t)]$, rather than
exponential \cite{JPB2}. Boyanovsky, de Vega, et al found the same behavior
in scalar QED \cite{DB1,DB2}.

\paragraph*{4. HTL transverse propagator with magnetic mass.}
For QCD the HTL propagators must be modified so as to include a magnetic mass
$\mu^{2}$. This does not change the asymptotic behavior of the pole
contribution. As shown In Sec. V A it does change the cut contribution by
eliminating the term of order
$T/r$ and modifying the coefficients of the $1/r^{3}$ terms. Using Eqs.
(\ref{429})
and (\ref{430}) gives
\begin{eqnarray}
^{*}\tilde{{\cal D}}^{ij\,\rm cut}(t,\vec{r})\to
-&&i{\tilde{b}_{t}^{\rm
cut}(\overline{x})\over t^{3}}
(-\delta^{ij}+3\hat{r}^{i}\hat{r}^{j})\nonumber\\
-&&i{\tilde{b}_{t}^{\prime\rm
cut}(\overline{x})\over t^{3}} (\delta^{ij}-\hat{r}^{i}\hat{r}^{j}).
\label{64}\end{eqnarray}
The coefficients depend on the magnetic mass through the denominator of the
transverse spectral function $\tilde{N}_{t}$ given in Eq. (\ref{428}):
\begin{eqnarray}
\tilde{b}_{t}^{\rm cut}(\overline{x})=&& {2T\over 3\pi
m_{g}^{2}\overline{x}^{3}}
\bigg\{
\int_{0}^{\overline{x}}\!dy{1-y^{2}
\over \tilde{N}_{t}(y,0)}
 -{\overline{x}(1-\overline{x}^{2})\over
\tilde{N}_{t}(\overline{x},0)}\bigg\},\nonumber\\
\tilde{b}_{t}^{\prime\rm cut}(\overline{x})=&&
{-2T\over 3\pi m_{g}^{2}\overline{x}}\,
{\partial\over\partial \overline{x}}\bigg[{1-\overline{x}^{2}
\over \tilde{N}_{t}(\overline{x},0)}\bigg].\nonumber
\end{eqnarray}
The more rapid fall-off in Eq. (\ref{64}) than for QED implies that the
quark damping rates are infrared finite \cite{sum}.

\subsection{Asymptotic space-like behavior of gauge boson propagators}

The deep space-like region requires  $t$ and $r$  both large with a fixed
ratio of $t/r$ satisfying $t/r<1$.  For the free propagator $r$ and $t$ must
be large compared to $1/2\pi T$; for the HTL propagator they must be large
compared to
$1/m_{g}$.
The leading term in the asymptotic space-like behavior is displayed in the
second
column of Table I.

\paragraph*{1. Free, massless gauge boson.} In the Coulomb gauge ${\cal
D}^{00}(t,r)=\delta(t)/4\pi r$. The asymptotic behavior of the transverse
propagator in the  deep space-like region can be found in \cite{AW}
\begin{mathletters}\begin{eqnarray}
{\cal D}^{00}(t,r)=&&0\\
{\cal D}^{ij}(t,\vec{r})\to &&{-iT\over
8\pi r}(\delta^{ij}+\hat{r}^{i}\hat{r}^{j})\nonumber\\
+&&{iT\over 8\pi r\overline{x}^{2}}(-\delta^{ij}+3\hat{r}^{i}\hat{r}^{j}).
\end{eqnarray}\label{65}\end{mathletters}
There are omitted terms of order $1/r^{2}$ and $1/r^{3}$.

\paragraph*{2. HTL longitudinal propagator.}
The asymptotic space-like behavior of the longitudinal gauge propagator
computed
in Sec. IV B can be summarized as follows:
\begin{mathletters}\begin{eqnarray}
^{*}{\cal D}^{00\,\rm pole}(t,r)\to &&-i{m_{g}c(t)\over8\pi  r}\label{66a}\\
 ^{*}{\cal D}^{00\,\rm cut}(t,r)\to&&{\rm exp.\; small}\end{eqnarray}
\label{66}\end{mathletters}
where
\begin{equation}
c(t)={e^{-im_{g}t}\over 1-e^{-\beta m_{g}}}
-{e^{im_{g}t}\over 1-e^{\beta m_{g}}}\end{equation}
The cut contribution is exponentially small as $r\to\infty$.
In Appendix \ref{sec:Euclidean} the same results are obtained in the
imaginary time
formalism. From that viewpoint, the static $n=0$ mode is exponentially small
because of Debye screening. The non-static contributions give exactly the same
asymptotic behavior as the pole contribution in Eq. (\ref{66a}).

\paragraph*{3. HTL transverse propagator.} From Sec. IV B the
asymptotic behavior of the transverse propagator in the deep space-like
region is given by
\begin{mathletters}\begin{eqnarray}
^{*}{\cal D}^{ij\,\rm pole}(t,\vec{r})\to&&
{-ic(t)\over 8\pi m_{g} r^{3}}(-\delta^{ij}+3\hat{r}^{i}\hat{r}^{j})\\
^{*}{\cal D}^{ij\,\rm cut}(t,\vec{r})\to&&
{-iT\over 8\pi r}(\delta^{ij}+\hat{x}^{i}\hat{x}^{j})\nonumber\\
+&&{iT\over 4\pi m_{g}^{2} r^{3}}(-\delta^{ij}+3\hat{r}^{i}\hat{r}^{j}).
\end{eqnarray}\label{68}\end{mathletters}
It is rather unexpected that the leading term $T/r$ comes entirely from the
branch
cut in the transverse propagator. The subleading term $1/(Tr^{3})$ comes both
from the  branch cut and from the pole. In Appendix \ref{sec:Euclidean} both
the leading and the subleading terms are obtained in the imaginary time
formalism. In that approach the leading $T/r$ term comes from the static
$n=0$ mode. The subleading
  contribution comes from summing over the nonstatic modes.
This applies to high temperature QED, where there is no magnetic mass.

\paragraph*{4. HTL transverse propagator with a magnetic mass.} For QCD
 a magnetic mass $\mu^{2}$ must be included. The asymptotic behavior of the
pole
contribution does not change. As shown in Sec. V B, the term of order $T/r$
in the
cut contribution is absent and the asymptotic behavior becomes
\begin{mathletters}\begin{eqnarray}
^{*}\tilde{{\cal D}}^{ij\,\rm pole}(t,\vec{r})\to&&
{-ic(t)\over 8\pi m_{g} r^{3}}(-\delta^{ij}+3\hat{r}^{i}\hat{r}^{j})\\
^{*}\tilde{{\cal D}}^{ij\,\rm cut}(t,\vec{r})\to &&
{iT\over 4\pi r^{3}}\bigg[{1\over m_{g}^{2}}
\!-\!{1\over \mu^{2}}\bigg](-\delta^{ij}+3\hat{r}^{i}\hat{r}^{j}).
\end{eqnarray}\label{69}\end{mathletters}
Exactly the same asymptotic behavior is obtained using the imaginary time
formalism
in Appendix \ref{sec:Euclidean}. In that approach the $1/\mu^{2}$ term
comes from the static $n=0$ mode and the other terms  come from
the nonstatic modes.

\subsection{Light-cone behavior}

The asymptotic behaviors computed
for $t/r>1$ and for $t/r<1$ do not allow any conclusion about the
behavior of the HTL propagators on the light cone, $t/r=1$.
To deduce the light-cone behavior, consider the inverse Fourier integral over
space-time:
\begin{equation}
^{*}\!D^{\mu\nu}_{>}(k_{0},\vec{k})=\int\!d^{4}x\,e^{iK\cdot x}
\;^{*}\!{\cal D}^{\mu\nu}_{>}(t,\vec{r}).
\end{equation}
After performing the angular integrations,  the phase can be expressed as
\begin{displaymath}
k_{0}t-kr={1\over 2}(k_{0}\!+\!k)(t\!-\!r)+{1\over 2}(k_{0}\!-\!k)(t\!+\!r).
\end{displaymath}
In the  limit $k_{0}+k\to\infty$ with
$k_{0}-k$ fixed
the integration  is dominated by the light cone  $t-r\to 0$.
 At large energy and momentum  the
 HTL propagator reduces to the free thermal propagator in the
Coulomb gauge \cite{MLB}. Thus
\begin{displaymath}
\lim_{k_{0}+k\to\infty}\,^{*}\!D^{ij}_{>}(k_{0},\vec{k})
=-i{2\pi\epsilon(k_{0})
\delta(k_{0}^{2}\!-\!k^{2})\over 1-e^{-\beta
k_{0}}}(\delta^{ij}-\hat{k}^{i}\hat{k}^{j}).\end{displaymath}
Since $\beta k_{0}\to\infty$, the limit is actually  independent of
temperature.  Because the momentum-space limit $k_{0}+k\to\infty$ with
$k_{0}-k$ fixed is controlled by the light cone $t-r\to 0$ this implies that
the HTL propagator in space-time should have the same light-cone behavior as
the free propagator in space-time. Near the light cone the singular behavior
of the free Coulomb gauge propagator is
\begin{eqnarray}
t\to r:\hskip0.3cm {\cal D}^{ij}_{>}(t,\vec{r})\to
{i\over
8\pi^{2}r(t-r)}&&\big(\delta^{ij}-\hat{r}^{i}\hat{r}^{j}\big)\nonumber\\
-{i\over
8\pi^{2}r}\ln(t-r)&&\big(\!-\delta^{ij}\!+\!3\hat{r}^{i}\hat{r}^{j}\big)
+\dots,
\nonumber\end{eqnarray}
with terms that are finite at $t=r$  omitted.
Therefore the HTL propagator $^{*}\!{\cal D}^{ij}_{>}(t,\vec{r})$ in the
Coulomb gauge  should have
these same singularities at $t=r$.

\acknowledgements

This work was supported in part by the U.S. National Science Foundation under
grant PHY-9900609. A preliminary version was presented at a Brookhaven National
Laboratory workshop. At that presentation the contribution of the cut in the
transverse propagator to the asymptotic behavior in the deep time-like region
(Sec. V A) was incorrect. I am  grateful for suggestions from Dan Boyanovsky,
Hector de Vega, Rob Pisarski, and Larry Yaffe that eventually led to the
correct
 result.

\appendix

\section{Spectral function in other gauges}

All calculations in the paper are done in the Coulomb gauge and the
spectral function has the structure given in Eq. (\ref{spectral}). In any
other gauge
the various components of $\rho^{\mu\nu}(K)$ can also be expressed  in
terms of the
longitudinal and transverse spectral functions.

A particularly simple choice is the temporal gauge, $A^{0}=0$. There are  seven
vanishing components of the spectral function:
$\rho^{00}=\rho^{0j}=\rho^{j0}=0$. The only nonvanishing components are
\begin{displaymath}
\rho^{ij}(K)=(\delta^{ij}-\hat{k}^{i}\hat{k}^{j})\rho_{t}(K)+{k^{i}k^{j}\over
k_{0}^{2}}\rho_{\ell}(K).\end{displaymath}

In the Landau gauge the various components of the spectral function are
\begin{eqnarray}
\rho^{00}(K)=&&{k^{4}\over (K^{2})^{2}}\rho_{\ell}(K)\nonumber\\
\rho^{0j}(K)=&& {k_{0}k^{2}k^{j}\over(K^{2})^{2}}\rho_{\ell}(K)\nonumber\\
\rho^{ij}(K)=&&{k_{0}^{2}k^{i}k^{j}\over (K^{2})^{2}}\rho_{\ell}(K)+
(\delta^{ij}-\hat{k}^{i}\hat{k}^{j})\rho_{t}(K).\nonumber
\end{eqnarray}
In more general covariant gauges the spectral function has an additional term
$K^{\mu}K^{\nu}\delta'(K^{2})$ that must be added to the Landau gauge result.

\section{Lemma on Asymptotic Space-like Behaviors}\label{sec:lemma}

 In the deep space-like region, $r\to\infty$
 with a fixed  ratio $r/t>1$, there are no points of stationary phase
on the real $k$ axis. For the pole contributions  the stationary
phase condition
$d\omega/dk=r/t$ cannot be satisfied for real $k$. For the cut contributions
the stationary phase condition $k_{0}/k=r/t$ cannot be satisfied since
$k_{0}<k$ on the Landau cut.
In the complex $k$ plane there
can be points of stationary phase but they will give
contributions that fall exponentially  with $r$.  However, the
leading asymptotic behavior  will typically be be a power-law falloff
 that is not related to the stationary phase points.
The generic integral to be investigated is
 \begin{equation}
{\cal D}(t,r)=\int_{0}^{\infty}dk\,{\sin kr\over kr}f(t,k).
\label{B1}\end{equation}
The function $f(t,k)$ will always be an even function of $k$ and infinitely
differentiable on the real $k$ axis. The lemma will show that  asymptotic
behavior
is controlled by the endpoint $k=0$. Specifically in the deep space-like region
\begin{equation}
{\cal D}(t,r)\to {\pi\over 2r}f(t,0) +{\rm exp.\; small}.\label{B2}
\end{equation}
This is  the form in which the lemma is used in Sec. V B.
However in Sec. IV B the naturally occurring
integral is of the form
\begin{displaymath}
{\cal D}(t,r)={1\over r}\int_{0}^{\infty}dk\,k\sin
(kr) F(t,k).
\end{displaymath}
By renaming $F(t,k)=f(t,k)/k^{2}$ it is clear that the asymptotic behavior of
the  integral
 will be $c_{0}\pi/2r$ where $F(t,k)\to c_{0}/k^{2}$ as $k\to
0$. It is this second form that occurs naturally in Sec. IV B.

To isolate the $1/r$
contribution,  write the original integral Eq. (\ref{B1}) as
${\cal D}={\cal D}_{f}+{\cal D}_{g}$ where
\begin{mathletters}\begin{eqnarray}
{\cal D}_{f}(t,r)=&&\int_{0}^{\infty}\!dk\,{\sin(kr)\over kr}e^{-\alpha
k}\;f(t,0)\\
{\cal D}_{g}(t,r)=&&\int_{0}^{\infty}\!dk\,{\sin(kr)\over r}
\,g(t,k),\end{eqnarray}\end{mathletters}
with $g(t,k)$  defined by
\begin{equation}
g(t,k)={1\over k}\big[f(t,k)-e^{-\alpha k}
f(t,0)\big].\end{equation}
The parameter $\alpha$  exactly cancels
in the sum ${\cal D}_{f}+{\cal D}_{g}$.

(1) The integral ${\cal D}_{f}$  is elementary and gives
\begin{displaymath}
{\cal D}_{f}(t,r)=f(t,0)\,{1\over r}\tan^{-1}(r/\alpha).
\end{displaymath}
Since $\alpha$ is a fixed parameter and $r\to\infty$, one can expand
the inverse tangent  to get
\begin{equation}
{\cal D}_{f}(t,r)=f(t,0){1\over r}\Big[{\pi\over 2}-\sum_{\ell=0}^{\infty}
{(-1)^{\ell}\over 2\ell+1}\Big({\alpha\over r}\Big)^{2\ell+1}\Big].
\label{D1}\end{equation}

(2) To investigate  ${\cal D}_{g}$ it is useful to use repeated integration by
parts. First write  ${\cal D}_{g}$ as
\begin{displaymath}
{\cal D}_{g}(t,r)={-1\over r^{2}}\int_{0}^{\infty}\!dk\,{d\cos(kr)\over dk}
g(t,k).\end{displaymath}
An integration by parts  gives
\begin{displaymath}
{\cal D}_{g}(t,r)={1\over r^{2}}g(t,0)+{1\over
r^{2}}\int_{0}^{\infty}\!dk\,\cos(kr)\,{dg(t,k)\over dk}.
\end{displaymath}
There is no surface term from the upper limit since  $g(t,\infty)=0$.
Next write $\cos(kr)=r^{-1}d\sin(kr)/dk$ and integrate by parts.
Because of $\sin(kr)$ there will be no contribution at $k=0$:
\begin{displaymath}
{\cal D}_{g}(t,r)={1\over r^{2}}g(t,0)-{1\over r^{3}}\int_{0}^{\infty}\!dk
\,\sin(kr){d^{2}g(t,k)\over dk^{2}}.
\end{displaymath}
Continuing this process gives
\begin{eqnarray}
{\cal D}_{g}(t,r)=\sum_{\ell=0}^{n}&&{(-1)^{\ell}\over
r^{2\ell+2}}\bigg[{d^{2\ell}g\over dk^{2\ell}}\bigg]_{k=0}\nonumber\\
+&&{(-1)^{n}\over
r^{2n+3}}\int_{0}^{\infty}\!\!dk\,\sin(kr)\,{d^{2n+2}g\over dk^{2n+2}}
.\nonumber\end{eqnarray}
From the definition of $g(t,k)$ the even derivatives are
\begin{displaymath}
\bigg[{d^{2\ell}g\over dk^{2\ell}}\bigg]_{k=0}
={1\over 2\ell+1}\bigg\{\bigg[{d^{2\ell+1}f\over dk^{2\ell+1}}\bigg]_{k=0}
+\alpha^{2\ell+1}f(t,0)\bigg\}.\end{displaymath}
The first term vanishes since $f(t,k)$ is an even function of $k$.
Therefore
\begin{eqnarray}
{\cal D}_{g}(t,r)=&&f(t,0){1\over r}\sum_{\ell=0}^{n}{(-1)^{\ell}\over
2\ell+1}\Big({\alpha\over r}\Big)^{2\ell+1}\nonumber\\
+&&{(-1)^{n}\over
r^{2n+3}}\int_{0}^{\infty}\!\!dk\,\sin(kr)\,{d^{2n+2}g\over dk^{2n+2}}.
\end{eqnarray}

(3) Adding together  ${\cal D}_{f}$ and ${\cal D}_{g}$
results in a
 cancellation of the powers $r^{-2}, r^{-4},\dots, r^{-2n-2}$ and gives
\begin{eqnarray}
{\cal D}(t,r)=&&f(t,0){\pi\over 2r}
\!+\!{(-1)^{n}\over
r^{2n+3}}\!\int_{0}^{\infty}\!\!dk\sin(kr)\,{d^{2n+2}g\over
dk^{2n+2}}\nonumber\\
-&&f(t,0){1\over r}\sum_{\ell=n+1}^{\infty}{(-1)^{\ell}\over
2\ell+1}
\Big({\alpha\over r}\Big)^{2\ell+1}.\end{eqnarray} This formula is exact. The
right hand side is independent of the choice of $\alpha$ and of $n$.
(For example, one could set $\alpha=0$ and eliminate the summation.)  The first
correction to
$1/r$ is of order
$1/r^{2n+3}$. Since  $f(t,k)$ is infinitely differentiable, $n$ may be
chosen arbitrarily large. This proves that the corrections to $1/r$ fall
faster than any power of $1/r$, i.e. fall exponentially with $r$. The
asymptotic behavior is therefore as given in Eq. (\ref{B2}).
It is, of course, possible for $f(t,0)$ to vanish.
In that case the leading behavior falls exponentially with $r$.

\section{Asymptotic space-like behavior using Imaginary time}
\label{sec:Euclidean}

The analysis of the hard-thermal-loop propagator in the deep
space-like region showed that both the longitudinal and transverse
propagators fall like $1/r$. In the longitudinal case the coefficient of $1/r$
depends on the thermal gluon mass $m_{g}$ and on time.  This contribution
comes entirely from the longitudinal quasiparticle pole. For the
transverse propagator the
$1/r$ contribution is the same as for a free, thermal field theory.
Rather surprisingly, this simple contribution comes
entirely from the cut in the transverse propagator.

This Appendix computes the asymptotic space-like behavior of the longitudinal
and transverse propagators using the imaginary time formalism. The
results obtained are identical with Eqs. (\ref{66}), (\ref{68}), and
(\ref{69}).
The asymptotic behavior of the longitudinal propagator, which comes
from the quasiparticle pole in the real-time method, will come from all
the non-static modes in the imaginary-time method.
The asymptotic behavior of the transverse propagator, which comes
from the branch cut in the real-time method, will come from entirely
from the static mode of the imaginary-time method.

\subsection{Asymptotic space-like behavior of $\;^{*}{\cal D}^{00}(t,r)$}

In imaginary time the propagator for the $A^{0}$ component of the vector
potential in the Coulomb gauge is
\begin{equation}
^{*}{\cal D}_{E}^{00}(\tau,r)=iT\!\!\sum_{n=-\infty}^{\infty}\!\!\int
\!{d^{3}k\over (2\pi)^{3}\,}{e^{i\vec{k}\cdot\vec{r}-i\omega_{n}\tau}
\over k^{2}+\Pi_{\ell}(i\omega_{n},k)},
\end{equation}
where $\omega_{n}=2\pi nT$ and $-\beta\le\tau\le\beta$. The denominator
depends on
$\omega_{n}$ only through the self-energy.  Separate this into the static,
n=0, mode and the nonstatic modes
\begin{equation}
{\cal D}_{E}^{00}(\tau,r)=
{\cal D}_{n=0}^{00}(\tau,r)
+{\cal D}_{n\neq 0}^{00}(\tau,r).
\end{equation}
The static value, $\Pi_{\ell}(0,k)=3m_{g}^{2}$, produces the standard
Debye screening result:
\begin{displaymath}
{\cal D}_{n=0}^{00}(\tau,r)\!
=iT\!\int\!{d^{3}k\over (2\pi)^{3}}\,{e^{i\vec{k}\cdot\vec{r}}
\over k^{2}+3m_{g}^{2}}
={iT\over 4\pi r}e^{-m_{g}r\sqrt{3}}.
\end{displaymath}
This exponentially small contribution is not the dominant one.

The nonstatic terms will turn out to dominate at large separations.
The reason is that for $n\neq 0$
the propagator denominator  vanishes at small $k$:
\begin{displaymath}
k\to 0:\hskip0.3cmk^{2}+\Pi_{\ell}(i\omega_{n},k)\to
k^{2}\Big(1+{m_{g}^{2}\over
\omega_{n}^{2}}\Big)+{\cal O}(k^{4}).
\end{displaymath}
Therefore the leading nonstatic contribution in the deep space-like region is
\begin{displaymath}
{\cal D}_{n\neq 0}^{00}(\tau,r)\to iT\int{d^{3}k\over
(2\pi)^{3}}{e^{-\vec{k}\cdot\vec{r}}\over k^{2}}
\sum_{n=-\infty}^{\infty}{\omega_{n}^{2}\over \omega_{n}^{2}+m_{g}^{2}}
e^{-i\omega_{n}\tau}.
\end{displaymath}
The integration gives
 $1/(4\pi r)$ and the sum is elementary. The sum converges for $\tau$ real
and gives
\begin{displaymath}
{\cal D}_{n\neq 0}^{00}(\tau,r)\to
{-im_{g}\over 8\pi r}\Big[{e^{-m_{g}\tau}\over 1-e^{-\beta m_{g}}}
+{e^{m_{g}\tau}\over e^{\beta m_{g}}-1}\Big].
\end{displaymath}
This can be continued from Euclidean time $\tau$ to real time $t$ by the
replacement
$\tau\to it$. Thus use the function
\begin{equation}
c(t)={e^{-im_{g}t}
\over 1-e^{-\beta m_{g}}}+{e^{im_{g}t}\over e^{\beta m_{g}}-1},
\end{equation}
that was introduced in Eq. (\ref{3B3}). The space-like asymptotic behavior
of the
Minkowski two-point function is
\begin{equation}
{\cal D}_{>}^{00}(t,r)\to
{-im_{g}c(t)\over 8\pi r}+{\rm exp.\; small}.
\end{equation}
This coincides with Eq. (\ref{66}). In the real-time calculation
this result comes entirely from the quasiparticle pole.

\subsection{Asymptotic space-like behavior of
 $^{*}{\cal D}^{\lowercase{ij}}(\lowercase{x})$}

The transverse propagator in Euclidean time is
\begin{displaymath}
^{*}\!{\cal D}_{E}^{ij}(\tau,\vec{r})
=-iT\!\!\sum_{n=-\infty}^{\infty}\!
\int\!\!{d^{3}k\over
(2\pi)^{3}}\,{e^{i\vec{k}\cdot\vec{r}-i\omega_{n}\tau}
(\delta^{ij}-\hat{k}^{i}\hat{k}^{j})
\over \omega_{n}^{2}+k^{2}+\Pi_{t}(i\omega_{n},k)}.
\end{displaymath}
As before,  separate out the static, $n=0$ mode from the
others:
\begin{displaymath}
^{*}\!{\cal D}_{E}^{ij}=^{*}\!{\cal D}_{n=0}^{ij}+^{*}\!{\cal D}_{n\neq
0}^{ij}.
\end{displaymath}
Because the transverse self-energy vanishes at $n=0$
the static propagator is the same as the free propagator:
\begin{eqnarray}
^{*}\!{\cal D}_{n=0}^{ij}(\vec{r})=&&
-iT\!\!
\int\!\!{d^{3}k\over
(2\pi)^{3}}\,{e^{i\vec{k}\cdot\vec{r}}
\over k^{2}}(\delta^{ij}-\hat{k}^{i}\hat{k}^{j})\nonumber\\
=&&{-iT\over 8\pi r}(\delta^{ij}+\hat{x}^{i}\hat{x}^{j}).\label{a4}
\end{eqnarray}
This will turn out to be the dominant term. It coincides with the
cut contribution  in Eq. (\ref{68}).

The remaining sum over non-zero modes can be written
\begin{displaymath}
^{*}\!{\cal D}_{n\neq 0}^{ij}(\tau,\vec{r})
=(-\delta^{ij}\nabla^{2}+\nabla^{i}\nabla^{j})
W_{n\neq 0}(\tau,r),
\end{displaymath}
where the new function is
\begin{displaymath}
W_{n\neq 0}(\tau,r)=-iT\!\sum_{n\neq 0}
\int\!{d^{3}k\over
(2\pi)^{3}}{e^{i\vec{k}\cdot\vec{r}-i\omega_{n}\tau}
\over k^{2}\big[\omega_{n}^{2}+k^{2}+\Pi_{t}(i\omega_{n},k)\big]}
\end{displaymath}
The dominant contribution at large $r$ comes from $k\to 0$.
In this limit $\Pi_{t}(i\omega_{n},0)=m_{g}^{2}$ and thus
\begin{eqnarray}
W_{n\neq 0}(\tau,r)\to &&-iT\!\sum_{n\neq 0}
\int\!{d^{3}k\over
(2\pi)^{3}}{e^{i\vec{k}\cdot\vec{r}-i\omega_{n}\tau}
\over k^{2}\big[\omega_{n}^{2}+m_{g}^{2}\big]}\nonumber\\
=&&{-iT\over 4\pi r}\sum_{n\neq 0}{e^{-i\omega_{n}\tau}
\over \omega_{n}^{2}+m_{g}^{2}}.\nonumber
\end{eqnarray}
The  sum is convergent for real $\tau$ and gives
\begin{displaymath}
W_{n\neq 0}(\tau,r)\!\to\!
{-iT\over 4\pi r}\bigg[\!{-1\over m_{g}^{2}}
+{1\over 2m_{g}T}\!\Big({e^{-m_{g}\tau}\over 1\!-\!e^{-\beta m_{g}}}
-{e^{m_{g}\tau}\over 1\!-\!e^{\beta m_{g}}}\Big)\!\!\bigg].
\end{displaymath}
The continuation from Euclidean $\tau$ to real $t$ requires the
replacement $\tau\to it$.  so that
\begin{displaymath}
W_{n\neq 0}(t,r)\to {iT\over 4\pi r}\bigg[{1\over m_{g}^{2}}-
{c(t)\over 2m_{g}T}\bigg]
\end{displaymath}
Performing the spatial derivatives gives for the asymptotic behavior of the
non-static propagator
\begin{equation}
^{*}{\cal D}^{ij}_{n\neq 0}(t,r)\to {iT\over 4\pi r^{3}}
\bigg[{1\over m_{g}^{2}}-
{c(t)\over 2m_{g}T}\bigg](-\delta^{ij}+3\hat{r}^{i}\hat{r}^{j}).
\label{Ag}
\end{equation}
The leading contribution is the static term,  Eq. (\ref{a4}).
This sub-leading term agrees with Eq. (\ref{68}). In the real-time
calculation the
sub-leading term comes both from the quasiparticle pole and from the branch
cut.

\paragraph*{Inclusion of a magnetic mass.} A magnetic mass eliminates the
$1/r$ contribution of the static mode but introduces a new $1/r^{3}$
static contribution. To see this, define
\begin{displaymath}
W_{n=0}(r)=-iT\!\int{d^{3}k\over
(2\pi)^{3}}{e^{i\vec{k}\cdot\vec{r}}
\over k^{2}\big[k^{2}+\Pi_{t}(0,k)\big]}.
\end{displaymath}
The angular integrations give
\begin{displaymath}
W_{n=0}(r)={-iT\over 2\pi^{2}}\!\int_{0}^{\infty}\!dk
\;{\sin kr\over kr}{1
\over k^{2}+\Pi_{t}(0,k)}.
\end{displaymath}
As $r\to\infty$ this integral is
dominated by the region $k\to 0$. One expects that  $\Pi_{t}(0,k)\to
\mu^{2}+ k^{2}$, in the limit  $k\to 0$.
 The asymptotic behavior is
\begin{eqnarray}
W_{n=0}(r)\to&& {-iT\over 2\pi^{2}}\!\int_{0}^{\infty}\!dk
\;{\sin kr\over kr}{1
\over k^{2}+\mu^{2}}\nonumber\\
\to &&{-iT\over 4\pi r\mu^{2}}+{\rm exp.\; small.}\nonumber
\end{eqnarray}
The exponential fall-off is of the form $\exp(-\mu r)$.
Differentiating this gives for the static part of the transverse propagator
\begin{displaymath}
^{*}\tilde{{\cal D}}^{ij}_{n=0}(r)\to {-iT\over 4\pi r^{3}\mu^{2}}
(-\delta^{ij}+3\hat{r}^{i}\hat{r}^{j}).
\end{displaymath}
Adding the non-static contribution from  Eq. (\ref{Ag}) gives
\begin{equation}
^{*}\tilde{{\cal D}}^{ij}(t,r)\to {iT\over 4\pi r^{3}}
\bigg[\!-\!{1\over \mu^{2}}\!+\!{1\over m_{g}^{2}}\!-\!{c(t)\over
2m_{g}T}\bigg]
(-\delta^{ij}\!+\!3\hat{r}^{i}\hat{r}^{j}).
\end{equation}
This agrees fully with the real-time result in Eq. (\ref{69}).

\section{Asymptotic time-like behavior of $^{*}{\cal D}^{\lowercase {ij\,\rm
cut}}_{>} \lowercase{(t,\vec{r})}$ }\label{sec:cut}

The tensor structure of the transverse Wightman function is
\begin{eqnarray}
^{*}{\cal
D}_{>}^{ij}(t,\vec{r})=\big(-\delta^{ij}+3\hat{r}^{i}\hat{r}^{j}\big)&&\,G(t,r)\
nonumber\\
+\big(\delta^{ij}-\hat{r}^{i}\hat{r}^{j}\big)\, &&H(t,r).
\end{eqnarray}
The contribution  of the Landau branch cut is particularly delicate to
calculate. In Sec. V A the cut contribution was split into a sum of two
parts:
\begin{eqnarray}
G^{\rm cut}(t,r)=&&G^{(1)}(t,r)+G^{(2)}(t,r),\nonumber\\
H^{\rm cut}(t,r)=&&H^{(1)}(t,r)+H^{(2)}(t,r),\nonumber\end{eqnarray}
in which $G^{(1)}$ and $H^{(1)}$ were of order $1/r$ plus order
$1/r^{3}$, whereas $G^{(2)}$ and $H^{(2)}$ were entirely of order
$1/r^{3}$.
This appendix contains the actual calculations of $H^{(1)}$ and
$G^{(2)}$ that were not presented in Sec. V A.

\subsection{Asymptotic behavior of $H^{(1)}(t,r)$}

The leading behavior of the cut contribution comes from the
integral $H^{(1)}$, defined in Sec. V A as
\begin{displaymath}
H^{(1)}(t,r)\!={-1\over
r(2\pi)^{2}}\int_{-\infty}^{\infty}\!dk\int_{-1}^{1}\!dx
\;e^{ik(r-xt)}\,I^{(1)}(x,k),
\end{displaymath}
with $I^{(1)}$  given in Eq. (\ref{4Ai1}).
To compute $H^{(1)}$ it is convenient to split the integration over $x$ into
three parts,
\begin{equation}
H^{(1)}(t,r)={-1\over r(2\pi)^{2}}\big(A+B
+C\big),\label{D2}
\end{equation}
with $A, B, C$ defined as
\begin{eqnarray}
A=&&\int_{-\infty}^{\infty}\!dk\int_{0}^{\overline{x}}\!dx
\,e^{ik(r-xt)}\;I^{(1)}(x,k),\nonumber\\
B=&&\int_{-\infty}^{\infty}\!dk\int_{\overline{x}}^{1}\!dx
\,e^{ik(r-xt)}\;I^{(1)}(x,k),\nonumber\\
C=&&\int_{-\infty}^{\infty}\!dk\int_{0}^{1}\!dx
\,e^{ik(r+xt)}\;I^{(1)}(x,k).\nonumber
\end{eqnarray}
At $\overline{x}=r/t$  the phase changes sign.
The $k$ integrations will be performed using Cauchy's theorem, since
the only singularities in $k$ of the integrand are simple poles.
The location of the
four poles are
\begin{displaymath}
k_{j}=e^{i\pi(2j-1)/4}m_{g}\Big({3\pi x\over 4}\Big)^{1/2},
\end{displaymath}
where $k_{1}, k_{2}, k_{3}, k_{4}$ are located in the first, second, third, and
fourth quadrants, respectively. The function $I^{(1)}$ can be written
  in terms
of partial fractions as
\begin{displaymath}
I^{(1)}(x,k)={3m_{g}^{2}T\over 16}\sum_{j=1}^{4}{1\over (k-k_{j})k_{j}^{2}}.
\end{displaymath}

(A) In integral $A$, the quantity
$r-xt$ is positive so that the integration over $k$ can be computed by adding a
semicircle of infinite radius in the upper-half of the complex $k$ plane.
Evaluating the $k$ integration by Cauchy's theorem  gives
\begin{displaymath}
A={T\over 2}\int_{0}^{\overline{x}}\! {dx\over x}\bigg[e^{ik_{1}(r-xt)}
-e^{ik_{2}(r-xt)}\bigg]
\end{displaymath}
Using  the values of $k_{1}$ and $k_{2}$ converts this to
\begin{displaymath}
A= iT\int_{0}^{\overline{x}}\! {dx\over x}
e^{-m\sqrt{x}(r-xt)}\sin\big[m\sqrt{x}(r-xt)\big],
\end{displaymath}
where $m=m_{g}\sqrt{3\pi/8}$.
For $r\to\infty$ and $t\to\infty$ with $r/t<1$, the integrand falls
exponentially
except at the two endpoints $x=0$ and $x=\overline{x}$. The contribution of
these
endpoints will be called $A_{0}$ and $A_{\overline{x}}$ and will fall like
a power of
$r$. Thus with exponentially small contributions omitted, one can say
$A=A_{0}+A_{\overline{x}}$.

To compute $A_{0}$ put $x=\varepsilon^{2}z^{2}$ where
\begin{equation}
\varepsilon={1\over ar}={1\over m_{g}r}\Big({8\over 3\pi}\Big)^{1/2}.
\end{equation}
The the contribution of the lower limit is
\begin{displaymath}
A_{0}=2iT\int_{0}^{z_{\rm max}}\! {dz\over z}\,e^{-z+\varepsilon^{2}
z^{3}/\overline{x}}
\sin\big[z-\varepsilon^{2} z^{3}/\overline{x}\big].
\end{displaymath}
 Expanding the
integrand to order  $\varepsilon^{2}$ and integrating gives
\begin{equation}
A_{0}=iT\Big[{\pi\over 2}+{2\varepsilon^{2}\over\overline{x}}
+{\cal O}(\varepsilon^{4})\Big].
\end{equation}
To compute the contribution of the upper endpoint, $x=\overline{x}$, return
to the
exact expression for $A$ and change variables to
$x=\overline{x}(1-\varepsilon z)$.
Then
\begin{displaymath}
A_{\overline{x}}=iT\varepsilon\int_{0}^{z_{\rm max}}\! {dz\over 1-\varepsilon
z}\,e^{-z\sqrt{\overline{x}(1-\varepsilon z)}}
\sin\Big[z\sqrt{\overline{x}(1-\varepsilon z)}\Big].
\end{displaymath}
Expanding the integrand to order $\varepsilon^{2}$ and integrating gives
\begin{equation}
A_{\overline{x}}=iT\Big[{\varepsilon\over 2\overline{x}^{1/2}}
+{\varepsilon^{2}\over\overline{x}}+{\cal O}(\varepsilon^{3})\Big].
\end{equation}

(B) In integral $B$, the quantity $r-xt$ is always negative. Therefore the $k$
contour can be closed in the lower half-plane and evaluated by Cauchy's
theorem.
The poles at $k_{3}$ and $k_{4}$ contribute and give
\begin{displaymath}
B=-{T\over 2}\int_{\overline{x}}^{1}\! {dx\over x}\bigg[e^{ik_{3}(r-xt)}
-e^{ik_{4}(r-xt)}\bigg].
\end{displaymath}
Substituting the value of the roots $k_{3}$ and
$k_{4}$ gives
\begin{displaymath}
B= iT\int_{\overline{x}}^{1}\! {dx\over x}
e^{m\sqrt{x}(r-xt)}\sin\big[m\sqrt{x}(r-xt)\big]
\end{displaymath}
Note that $r-xt\le 0$ in this integral. The integrand is exponentially
suppressed
except at the lower endpoint $x=\overline{x}$. This contribution
$B_{\overline{x}}$
will fall like a power of $r$. To compute this, put
$x=\overline{x}(1+\varepsilon z)$
so that
\begin{displaymath}
B_{\overline{x}}=-iT\varepsilon \int_{0}^{z_{\rm max}}\! {dz\over 1+\varepsilon
z}\,e^{-z(1+\varepsilon z)^{1/2}}\sin\big[z(1+\varepsilon z)^{1/2}\big].
\end{displaymath}
Expanding this for small $\varepsilon$ and integrating gives
\begin{equation}
B_{\overline{x}}=iT\Big[-{\varepsilon\over
2\overline{x}^{1/2}}+{\varepsilon^{2}\over
\overline{x}}+{\cal O}(\varepsilon^{3})\Big].
\end{equation}

(C) In integral $C$, since $r+xt$ is always positive, the $k$ contour can be
closed in the upper-half plane and evaluated with Cauchy' theorem:
\begin{displaymath}
C={T\over 2}\int_{0}^{1}\! {dx\over x}\bigg[e^{ik_{1}(r+xt)}
-e^{ik_{2}(r+xt)}\bigg].
\end{displaymath}
Using  the values of $k_{1}$ and $k_{2}$ gives
\begin{displaymath}
C= iT\int_{0}^{1}\! {dx\over x}
e^{-m\sqrt{x}(r+xt)}\sin\big[m\sqrt{x}(r+xt)\big],
\end{displaymath}
where $m=m_{g}\sqrt{3\pi/8}$.  The integrand falls exponentially except at
the lower
endpoint $x=0$. To compute this contribution, put $x=\varepsilon^{2}z^{2}$ to
obtain
 \begin{displaymath}
C_{0}=2iT\int_{0}^{z_{\rm max}}\! {dz\over z}\,e^{-z-\varepsilon^{2}
z^{3}/\overline{x}}
\sin\big[z+\varepsilon^{2} z^{3}/\overline{x}\big].
\end{displaymath}
Expanding in $\varepsilon$ and integrating gives
\begin{equation}
C_{0}=iT\Big[{\pi\over 2}-2{\varepsilon^{2}\over \overline{x}}+{\cal
O}(\varepsilon^{4})\Big].
\end{equation}

\paragraph*{Total for $H^{(1)}$.} The sum of the endpoint contributions from
$x\approx 0$ is
\begin{equation}
A_{0}+C_{0}=iT\Big[\pi+{\cal O}(\varepsilon^{4})
\Big].\end{equation}
The contributions from $x\approx \overline{x}$ sum to
\begin{equation}
A_{\overline{x}}+B_{\overline{x}}=iT\Big[2{\varepsilon^{2}\over\overline{x}}
+{\cal O}(\varepsilon^{3})\Big].
\end{equation}
It is quite likely that the ${\cal O}(\varepsilon^{3})$ terms actually cancel
 in this
sum, but it will not matter below. Using these in Eq. (\ref{D2}) gives for
the asymptotic behavior of
$H^{(1)}$:
\begin{equation}
H^{(1)}(t,r)\to {-iT\over 4\pi r}-{4iT\over 3\pi^{3}m_{g}^{2}r^{3}\overline{x}}
+{\cal O}\Big({1\over r^{4}}\Big).
\end{equation}
This value is used in Eq. (\ref{4H1}).

\subsection{Asymptotic behavior of $G^{(2)}(t,r)$ }

In Sec. V A the asymptotic value of $H^{(2)}$ was computed by the method of
stationary phase. This did not completely determine $G^{(2)}$ and so it too
must be computed by the method of stationary phase.

It is convenient to start with the original definition of $G$ in Eq.
(\ref{411}). If the angular integration over $z=\hat{k}\cdot\hat{r}$ is not
performed, the cut contribution is
\begin{equation}
G^{\rm cut}(t,r)={-i\over 16\pi^{2}}\int_{-1}^{1}\! dz(1-z^{2})\,
{\cal G}^{\rm cut}(t,r,z),\label{D11}
\end{equation}
where
\begin{displaymath}
{\cal G}^{\rm cut}(t,r,z)=\int_{-\infty}^{\infty}\! dk\int_{-1}^{1}\!dx
\,e^{ik(rz-xt)}\, kI_{t}(x,k),
\end{displaymath}
with $I_{t}$ defined in Eq. (\ref{4AIt}). If the $z$ integration is
performed at this stage it will change $kI_{t}$ to $I_{t}/k$ and therefore
prevent convergence at $k=0$. Separating $I_{t}=I^{(1)}+I^{(2)}$ as
given in Eq. (\ref{4Ai2}) gives
\begin{displaymath}
{\cal G}^{(2)}(t,r,z)=\int_{-\infty}^{\infty}\! dk\int_{-1}^{1}\!dx
\,e^{ik(rz-xt)}\, kI^{(2)}(x,k).
\end{displaymath}
The integrand $kI_{2}$ vanishes like $k^{2}$ at small $k$. The
stationary phase point is $\overline{\overline{x}}=rz/t$, $\overline{k}=0$.
Making the variable change analogous to Eq. (\ref{uv}) gives
\begin{displaymath}
{\cal G}^{(2)}(t,r,z)={2\over t}\int_{0}^{u_{\rm max}}\! udu\int_{0}^{2\pi}\!
d\theta\,e^{-u^{2}}\, kI^{(2)}(x,k).
\end{displaymath}
The Gaussian integration can be performed using Eq. (\ref{3A6}) to obtain
the asymptotic behavior
\begin{displaymath}
{\cal G}^{(2)}(t,r,z)\to {-\pi\over t^{3}}\bigg[{\partial^{4}
\big[kI^{(2)}(x,k)\big]\over \partial x^{2}\partial
k^{2}}\bigg]_{\overline{\overline{x}},\overline{k}}.
\end{displaymath}
This simplifies to
\begin{equation}
{\cal G}^{(2)}(t,r,z)\to{-8\pi T\over 3m_{g}^{2}t^{3}}
{\partial^{2}\over\partial \overline{\overline{x}}^{2}}
\bigg[{1-\overline{\overline{x}}^{2}\over N_{t}(\overline{\overline{x}},0)}
-{1\over(\pi\overline{\overline{x}})^{2}}\bigg].
\end{equation}
The stationary point depends on the angular variable $z$ since
$\overline{\overline{x}}=\overline{x}z$ with $\overline{x}=r/t<1$.
When ${\cal G}^{(2)}$ is substituted into Eq. (\ref{D11}) the integration
over $z$ must be performed. It is convenient to change integration variables
from $z$ to $y=\overline{x}z$ so that
\begin{displaymath}
G^{(2)}(t,r)={-i\over 16\pi^{2}}\int_{-\overline{x}}^{\overline{x}}
\!{dy\over\overline{x}}\Big(1-{y^{2}\over \overline{x}^{2}}\Big){\cal
G}^{2}(t,r,z).
\end{displaymath}
Using the asymptotic form for ${\cal G}^{(2)}$ gives
\begin{eqnarray}
G^{(2)}(t,r)\to {iT\over 3\pi m_{g}^{2} r^{3}}&&\int_{0}^{\overline{x}}
\!dy \,(\overline{x}^{2}-y^{2})\nonumber\\
&&\times{\partial^{2}\over\partial y^{2}}
\bigg[{1-y^{2}\over N_{t}(y,0)}-{1\over (\pi y)^{2}}\bigg].
\end{eqnarray}
The quantity in square brackets is an even function of $y$ that is finite at
$y=0$. The first $y$ derivative vanishes at $y=0$. Therefore one can
integrate by parts to obtain
\begin{eqnarray}
G^{(2)}(t,r)\to {2iT\over 3\pi m_{g}^{2} r^{3}}\int_{0}^{\overline{x}}
\!dy \,y
{\partial\over\partial y}
\bigg[{1-y^{2}\over N_{t}(y,0)}-{1\over (\pi y)^{2}}\bigg].
\nonumber\end{eqnarray}
This is the result used in Eq. (\ref{4G2}).


\references


\bibitem{P1}  R.D. Pisarski, Nucl. Phys. {\bf B309}, 476 (1988); {\bf
A498}, 423c (1989); Physica A {\bf 158}, 146 (1989).

\bibitem{P2} R.D. Pisarski, Phys. Rev. Lett. {\bf 63}, 129 (1989).


\bibitem{BP1} E. Braaten and R.D. Pisarski, Nucl. Phys. {\bf B337}, 569 (1990)
and {\bf B339}, 310 (1990).

\bibitem{BP2} E. Braaten and R.D. Pisarski, Phys. Rev. Lett. {\bf 64}, 1338
(1990);
Phys. Rev. D {\bf 42}, 2156 (1990); {\bf 46}, 1829 (1992).

\bibitem{sum} R.D. Pisarski, Phys. Rev. D {\bf 47}, 5589 (1993).


\bibitem{JPB} J.P. Blaizot and  E. Iancu, Phys. Rev. Lett. {\bf 70}, 3376
(1993); Nucl. Phys. {\bf B421}, 565 (1994); {\bf B434}, 662 (1995).


\bibitem{MLB} M. Le Bellac, {\it Thermal Field Theory} (Cambridge
University Press,
Cambridge, England, 1996).


\bibitem{HAW} H.A. Weldon, Phys. Rev. D {\bf 26}, 1394 (1982).

\bibitem{Mex} H.A. Weldon, in {\it First Latin American Symposium on High
Energy Physics and VII Mexican School  of Particles and Fields} ed. J.C.
D'Olivo, M. Klein-Kreisler, and H. Mendez (AIP COnference Proceedings {\bf
400}, Woodbury, N.Y., 1997), p. 424.

\bibitem{DB1} D. Boyanovsky, H.J. de Vega, R. Holman, S.P. Kumar, and R.D.
Pisarski, Phys. Rev. D. {\bf 58}, 125009 (1998).

\bibitem{DB2} D. Boyanovsky, H.J. de  Vega, R. Holman, and M. Simionato,
Phys. Rev. D. {\bf 50}, 065003 (1999).

\bibitem{Y1} P. Arnold and L.G. Yaffe, Phys. Rev. D {\bf 57}, 1178 (1998).

\bibitem{AW} H.A. Weldon, Phys. Rev. D {\bf 62}, 056003 and 056010 (2000).


\bibitem{cop} E.T. Copson, {\it Asymptotic Expansions} (Cambridge Univ.
Press, Cambridge, England, 1967).

\bibitem{din} R.B. Dingle, {\it Asymptotic Expansions: Their Derivation
and Interpretation} (Academic Press, London, England, 1973).


\bibitem{L} A.D. Linde, Phys. Lett. B {\bf 96}, 289 (1980).

\bibitem{GPY} D.J. Gross, R.D. Pisarski, and L.G. Yaffe, Rev. Mod. Phys.
{\bf 53} 43 (1981).

\bibitem{JPB2} J.P. Blaizot and E. Iancu, Phys. Rev. D {\bf 55}, 973 (1997);
{\bf 56}, 7877 (1997).
\end{document}